\documentclass[aps,showpacs,amssymb,amsfonts,twocolumn,pra]{revtex4}
\usepackage{amsmath,bbm,mathrsfs}
\usepackage[pdftex]{graphicx}
\usepackage{amsfonts}
\usepackage{amssymb}

\pdfoutput=1

\numberwithin{equation}{section}

\def\textbf#1{{\bf #1}}
\def\be{\begin{equation}}
\def\ee{\end{equation}}
\def\ben{\begin{eqnarray}}
\def\een{\end{eqnarray}}
\def\eea{\end{array}}
\def\bea{\begin{array}}
\newcommand{\ot}[0]{\otimes}
\newcommand{\Tr}[1]{\mathrm{Tr}#1}
\newcommand{\bei}{\begin{itemize}}
\newcommand{\eei}{\end{itemize}}
\newcommand{\ket}[1]{|#1\rangle}
\newcommand{\bra}[1]{\langle#1|}

\newcommand{\proj}[1]{\ket{#1}\bra{#1}}
\newcommand{\braket}[2]{\langle{#1}|{#2}\rangle}

\newcommand{\meanb}[1]{\big\langle\hspace{-0.1cm}\big\langle#1\big\rangle\hspace{-.1cm}\big\rangle_{\varrho}}

\def\blacksquare{\vrule height 4pt width 3pt depth2pt}

\begin{document}

\title{Beyond the standard entropic inequalities: stronger scalar separability criteria and
their applications.}

\author{Remigiusz Augusiak}
\email{remik@mif.pg.gda.pl}
\author{Julia Stasi\'nska}\email{jul.sta@gmail.com}
\author{Pawe\l{} Horodecki}\email{pawel@mif.pg.gda.pl}

\affiliation{Faculty of Applied Physics and Mathematics, Gda\'nsk
University of Technology, Narutowicza 11/12, 80-952 Gda\'nsk,
Poland}

\begin{abstract}
Recently it was shown that if a given state fulfils the reduction
criterion, it must also satisfy the known entropic inequalities.
The natural question arises as to whether it is possible to derive
some scalar inequalities stronger than the entropic ones, assuming
that stronger criteria based on positive but not completely
positive maps are satisfied. In the present paper we show that if
certain conditions hold, the extended reduction criterion [H.-P.
Breuer, Phys. Rev. Lett {\bf 97}, 080501 (2006); W. Hall, J. Phys.
A {\bf 40}, 6183 (2007)] leads to some entropic-like inequalities,
much stronger than their entropic counterparts. The comparison of
the derived inequalities with other separability criteria shows
that such an approach might lead to strong scalar criteria
detecting both distillable and bound entanglement. In particular,
in the case of $\mathrm{SO}(3)$-invariant states it is shown that
the present inequalities detect entanglement in regions, in which
linear entanglement witnesses based on the extended reduction map
fail. It should also be emphasized that in the case of $2\ot d$
states the derived inequalities detect entanglement efficiently,
while the extended reduction maps are useless, when acting on the
qubit subsystem. Moreover, there is a natural way to construct a
many-copy entanglement witnesses based on the derived inequalities
so, in principle, there is a possibility of experimental
realization. Some open problems and possibilities for further
research are outlined.
\end{abstract}


\maketitle

\section{Introduction.}
\setcounter{equation}{0}

Quantum entanglement, well understood for pure states
\cite{EPR,Sch}, was much later formalized for mixed states
\cite{Werner} and developed into a key ingredient of quantum
information theory, including especially quantum communication
(see Ref. \cite{RevModPhys}, and references therein). In the
bipartite case, a mixed quantum state acting on a
finite-dimensional Hilbert space ${\cal H}_{AB}=\mathcal{H}_{A}
\otimes \mathcal{H}_{B}$ is called separable if and only if it is
of the form \cite{Werner}
\begin{equation}\label{separable}
\varrho=\sum_{i}p_{i}\rho^{(i)}_{A}\ot\tilde{\rho}^{(i)}_{B}.
\end{equation}
Otherwise it is called entangled or inseparable. In the above
formula $\rho_{A}^{(i)}$ and $\tilde{\rho}_{B}^{(i)}$ are density
matrices acting on the Hilbert spaces $\mathcal{H}_{A}$ and
$\mathcal{H}_{B}$, respectively, $p_{i}\geq 0$ and
$\sum_{i}p_{i}=1$. The definition is consistent with the pure
state scenario, in which the state is entangled if and only if the
vector representing it
$\ket{\psi}\in\mathcal{H}_{A}\ot\mathcal{H}_{B}$ is not a tensor
product of vectors describing the subsystems
\begin{equation}
\ket{\psi} \neq \ket{\varphi} \otimes \ket{\phi},
\end{equation}
where $\ket{\varphi}\in\mathcal{H}_{A}$ and $\ket{\phi}\in\mathcal{H}_{B}$.

Schr\"odinger \cite{Sch} pointed out that the essence of pure
entangled state is of the informational kind, i.e., the total
information about the system exceeds the information about its
subsystems. In fact, the total information is maximal (since the
state is pure) while the local ones are not (since the subsystems
are mixed). For mixed states the above Schr\"odinger intuition was
first formalized in terms of the von Neumann entropy
$S_{1}(\varrho)= -\Tr(\varrho \log \varrho)$. Namely, it was
observed in Ref. \cite{RPHPLA} that any separable state has to
obey the converse rule, i.e., it must have the entropy of the
total system greater than entropies of the subsystems
\begin{equation}\label{vNentropy}
S_{1}(\varrho_{A})\leq S_{1}(\varrho)\quad \mathrm{and}\quad S_{1}(\varrho_{B}) \leq S_{1}(\varrho),
\end{equation}
where $\varrho_{A(B)}=\Tr_{B(A)}\varrho$. Thus any violation of
the above conditions {\it implies} entanglement (see also Ref.
\cite{Plastino} for an analysis of special examples). Recently
this fact was shown to play a central role in the quantum version
of Slepian-Wolf theorem \cite{Nature}, which solves the
long-standing open problem (analyzed first for pure states in Ref.
\cite{CA}) of full physical interpretation of negative quantum
conditional entropy $S_{1}(\varrho) - S_{1}(\varrho_{A})$. In
particular, it stimulated the development of operational approach
to other quantum conditional quantities \cite{DevetakYard}. Note
also that the conditional entropy of another kind, based on
$\alpha$-entropy with $\alpha=\infty$ (see below), happens to play
an important role in some cryptographic scenarios \cite{Renner}.

The condition (\ref{vNentropy}) belongs to the so-called scalar
criteria of entanglement. Its generalization, stating that any
separable state should satisfy
\begin{equation}\label{entropic}
S_{\alpha}(\varrho_{A})\leq
S_{\alpha}(\varrho)\quad\mathrm{and}\quad S_{\alpha}(\varrho_{B})
\leq S_{\alpha}(\varrho) \quad (\alpha\in[0,\infty))
\end{equation}
were derived first for special values of the parameter $\alpha$
\cite{RPHPLA,HHHEntr,CAG} and special class of separable states
\cite{MRHPRA}. Later Eq. (\ref{entropic}) was proved to hold for
the whole range of $\alpha \in [0,\infty)$
\cite{Terhal3,VollbrechtWolf}. Here, by $S_{\alpha}$ we denote,
e.g., the Renyi entropy defined as
\begin{equation}
S_{\alpha}^{R}=\frac{1}{1-\alpha}\log\Tr\varrho^{\alpha}.
\end{equation}
Straightforward calculations lead to more operational forms
of the inequalities (\ref{entropic}), which for $\alpha\in(1,\infty)$ become
\begin{equation}\label{entropic0}
\Tr\varrho_{A}^{\alpha}\geq \Tr\varrho^{\alpha}\quad\mathrm{and}\quad
\Tr\varrho_{B}^{\alpha}\geq \Tr\varrho^{\alpha},
\end{equation}
while for $\alpha\in[0,1)$,
\begin{equation}
\Tr\varrho_{A}^{\alpha}\leq \Tr\varrho^{\alpha}\quad\mathrm{and}\quad
\Tr\varrho_{B}^{\alpha}\leq \Tr\varrho^{\alpha}.
\end{equation}
Let us recall that for $\alpha=1$ the Renyi entropy reduces to the
von Neumann entropy $S_{1}$. For $\alpha=0$ we have
$S_{0}^{R}(\varrho)=R(\rho)$ with $R(\cdot)$ denoting the rank of
a given matrix. Finally, for $\alpha=\infty $,
$S^{R}_{\infty}(\varrho)=-\log||\varrho||$, where $||\cdot||$ is a
standard operator norm. Thus for $\alpha=\infty$ the conditions
(\ref{entropic}) become \cite{MHPH}:
\begin{equation}\label{entropic_niek}
||\varrho_{A}||\geq ||\varrho||\quad\mathrm{and}\quad
||\varrho_{B}||\geq ||\varrho||.
\end{equation}

It is worth mentioning that the above entropic criteria can be
viewed as a prototype of nonlinear separability criteria that has
been recently intensively developed in Refs.
\cite{nonlinear,Guhne}. In particular, the new class of entropic
inequalities that involve Klein-like entropies, i.e., entropies of
output statistics of measurements \cite{Guhne}.

Recently an experimental illustration of the inequality
(\ref{entropic0}) for $\alpha=2$ has been performed \cite{Bovino}.
For experimental realizations of other quantitative and
qualitative nonlinear separability tests see, e.g., Ref.
\cite{spinsqetal}.

Apart form the scalar criteria discussed above, the so called
structural criteria were introduced \cite{Peres,HHHPLA96} and
investigated (see Ref. \cite{maps} and references therein). Here
we shall especially need the separability conditions based on
positive but not completely positive maps \cite{HHHPLA96} (denoted
hereafter by $\Lambda$) with the positive partial transposition
(PPT) criterion \cite{Peres} as the most famous example. Positive
maps, characterizing separability themselves, allow also for
introduction of a dual picture, i.e., the description in terms of
the so-called {\it entanglement witnesses}
\cite{HHHPLA96,Terhal1,Terhal2}. Let us recall, that a Hermitian
operator $W$ is called an entanglement witness if its mean value
on all separable states is nonnegative and negative for at least
one entangled state. Entanglement witnesses lead to a popular
method of experimental entanglement detection nowadays (see Ref.
\cite{RevModPhys}). However, some other indirect applications of
the positive map criterion were also proposed. In particular, the
possible measurement of certain functionals of $\varrho$ and
$[I\ot\Lambda](\varrho)$ was discussed in Refs.
\cite{Carteret,PHRAMD}. Here and further by $I$ we shall be
denoting an identity map.

One of the criteria, based on positive maps and important from the
communication point of view is the so-called reduction criterion
\cite{reduction2,xor}. It arises from the reduction map, which
acts on a $d\times d $ matrix $A$ as $\Lambda_{r}(A)=(\Tr A
)\mathbbm{1}_{d} -A$. The criterion states that any separable
state $\varrho$ acting on
$\mathbb{C}^{d_{A}}\ot\mathbb{C}^{d_{B}}$, should retain a
nonnegative spectrum after the action of the map $I \otimes
\Lambda_{r}$, leading to the following operator inequality:
\begin{equation} \label{reduction}
\varrho_{A} \ot \mathbbm{1}_{d_{B}} \geq \varrho.
\end{equation}
In Ref. \cite{CAG} the above criterion was shown to imply the
first entropic inequality (\ref{vNentropy}). Later in Ref.
\cite{VollbrechtWolf} the implication was extended to all entropic
inequalities. In this way the criterion based on the positive map
provided the series of scalar criteria which for a natural number
$\alpha$ may be measured via the {\it collective entanglement
witnesses} (see, e.g., Refs. \cite{Bovino,PH3}).

In analogy to Refs. \cite{CAG,VollbrechtWolf} it is natural to ask a
general question. {\it Is it possible to derive entropic-like
inequalities from other positive maps than the reduction one?}

Recently, a new positive map, whose structure is similar to the
reduction map, has been introduced in Refs. \cite{Bcrit, Bcrit2,
Hall}. The map leads to the following operator inequality
\begin{equation}
\varrho_{A} \ot \mathbbm{1}_{d_{B}} \geq
\varrho+\varrho^{\tau_{B}^{U}},
\end{equation}
and unlike the reduction map, was shown to be indecomposable. As
such it can detect PPT entangled states \cite{BE}. Here
$\tau_{B}^{U}$ stands for partial transposition with respect to
subsystem $B$ composed with a local antisymmetric operation $U$ such
that $U^{\dagger}U\leq \mathbbm{1}_{d}$ on the second subsystem.
Of course, one may write a similar operator inequality for the
subsystem $A$. Using this map we give a partially positive answer to
the posed question.
For a large class of states satisfying additional assumptions
(including, in particular, the states that are isomorphic to
quantum channels) we derive a series of entropic-like inequalities
which detect entanglement better than their entropic counterparts.
We derive also the operator version of the inequalities.

The paper is organized as follows. The detailed construction of
the inequalities is given in Sec. \ref{sec1}. At the beginning we
discuss the case of two-particle states consisting of a qubit and
qudit (qubit-qudit states) to introduce the method and discuss
some special cases and examples. Then we present the inequalities
for higher-dimensional systems and give some illustrative
examples. In particular, we compare the derived inequalities with
the entropic inequalities and entanglement witness arising from
the Breuer criterion \cite{Bcrit}. In Sec. \ref{witness} we
present the corresponding multicopy entanglement witness. In Sec.
\ref{IV} we discuss in more details a special inequality which,
similarly to the entropic one for $\alpha=2$, can be measured as a
collective entanglement witness on two copies of a state. Finally,
using the fact that bipartite systems of even dimensions can be
simulated by multiqubit states we show in Secs. \ref{IV.B} and
\ref{IV.C} how to check the inequality experimentally within
coalescence-anticoalescence experimental setups known already from
the literature \cite{Bovino}.

\section{Inequalities}\label{sec1}
\setcounter{equation}{0}
The construction of entropic-like inequalities is based on the
recently introduced positive but not completely positive
indecomposable map \cite{Bcrit}, which acts on a $d\times d$
matrix $A$ (here $d$ is an even number) as follows:
\begin{equation}\label{Bmap}
\Lambda_B(A)=(\Tr A)\mathbbm{1}_{d}-A-A^{\tau}.
\end{equation}
The symbol ${\tau}$ denotes the time reversal of $A$, namely
$A^{\tau}=VA^{T}V^{\dagger}$, where $V$ is an antisymmetric
anti-diagonal unitary matrix with anti-diagonal elements $\pm 1$,
$\mathbbm{1}_{d}$ is a $d\times d$ identity matrix, and
superscript $T$ denotes the matrix transposition in the standard
basis. This map belongs to the class of indecomposable positive
maps $\Lambda_{U}^{(-)}$ introduced by Hall \cite{Hall}, where
instead of the particular $V$, an arbitrary antisymmetric
$(U^{T}=-U)$ matrix $U$ such that $U^{\dagger}U\leq
\mathbbm{1}_{d}$ is taken. The map can be written in the following
form
\begin{equation}\label{Hallmap}
\Lambda_{U}^{(-)}(A)=(\Tr A)\mathbbm{1}_{d}-A-UA^{T}U^{\dagger}.
\end{equation}
Note that for even $d$ one may take $U$ to be unitary since
only in this case antisymmetric unitaries exist \cite{Hall}.
In further considerations we will concentrate on the special case
considered by Breuer \cite{Bcrit}, however, throughout the paper
we will state the facts for general map $\Lambda_{U}^{(-)}$
whenever possible.

Let us also introduce a positive map similar to the Breuer-Hall
map that will become useful in further considerations. The only
difference a is a change of the sign before the modified
transposition map, i.e.,
\begin{equation}\label{phimap}
\Lambda_{U}^{(+)}(A)=(\Tr A)\mathbbm{1}_{d}-A+UA^{T}U^{\dagger},
\end{equation}
where A is again a $d\times d$ matrix. The proof of positivity
goes along the same lines as the proof for Breuer-Hall criterion
given in Refs. \cite{Bcrit,Hall}. Notice that
$\Lambda_{r}=(1/2)[\Lambda_{U}^{(+)}+\Lambda_{U}^{(-)}]$ and
$\tau^{U}=(1/2)[\Lambda_{U}^{(+)}-\Lambda_{U}^{(-)}]$.

Before we state the main results let us introduce the following notations:
\begin{equation}
X^{\tau^{U}}=UX^{T}U^{\dagger}=\tau^{U}(X),
\end{equation}
$\tau_{A}^{U}=\tau^{U}\ot I$, and, respectively,
$\tau_{B}^{U}=I\ot\tau^{U}$. As previously stated, in particular
case when $U=V$, the notation $\tau^{V}\equiv \tau$ shall be used.
Finally, we shall denote the standard partial transposition with
respect to the subsystem $A(B)$ by superscript $T_{A(B)}$, i.e.,
$[I\ot T](X)=X^{T_{B}}$ and $[T\ot I](X)=X^{T_{A}}$.
\subsection{The case of qubit-qudit}\label{sub-2.A}
As an introductory example we present the entropic-type
inequalities for the qubit-qudit states.
It should be emphasized that the Breuer map $\Lambda_{B}$ cannot
be used as a separability criterion in the case of the qubit-qudit
states (when the map acts on the smaller subsystem), since it
gives zero on arbitrary projector acting on $\mathbbm{C}^{2}$.
(Hall map $\Lambda_{U}^{(-)}$ is equivalent to Breuer map in this
case, since each unitary antisymmetric matrix $U$ acting on the
two dimensional subsystem can be written as $e^{i\phi}V$, which
does not change the map). However, it does not mean that it is not
useful in detecting entanglement at all. As we will see below it
is a good starting point for derivation of some inequalities.

Let us first recall the Hilbert-Schmidt form of any qubit-qudit
state. If we denote by $\varrho$ the density operator acting on
the Hilbert space $\mathbb{C}^{2}\ot\mathbb{C}^{d}$ and $\{f_i\}$
the generators of $\mathrm{SU}(d)$ with $f_{0}=\mathbbm{1}_{d}$,
then the density matrix $\varrho$ might be written in the product
basis $\{\sigma_{i}\ot f_{j}\}$ as
\begin{equation}\label{HS}
\varrho=\frac{1}{2d} \sum_{i=0}^{3}\sum_{j=0}^{d^{2}-1} \xi_{ij}
\sigma_i \ot f_j.
\end{equation}
On the first subsystem the basis is chosen to be Pauli matrices
defined as
\begin{equation}\label{Pauli}
\sigma_{1}=
\left(
\begin{array}{cc}
0 & 1\\
1 & 0
\end{array}
\right),\quad
\sigma_{2}=\left(
\begin{array}{cc}
0 & -i\\
i & 0
\end{array}
\right),\quad
\sigma_{3}=\left(
\begin{array}{cc}
1 & 0\\
0 & -1
\end{array}
\right)
\end{equation}
with $\sigma_{0}=\mathbbm{1}_{2}$. Coefficients $\xi_{ij}$ are
given by $\xi_{ij}=(d/2)\Tr(\varrho\sigma_{i}\ot f_{j})$,
$\xi_{00}=1$, $\xi_{i0}=\Tr \varrho_A\sigma_i$, $\xi_{0j}=\Tr
\varrho_B\sigma_j$, and thus real. The convention is such that
$\Tr f_i f_j=2\delta_{ij}$, for $i=1,\ldots,d$.

In the Hilbert-Schmidt formalism one may easily recognize how the
map $\tau_{A}$ acts on $\varrho$. When acting on the
two-dimensional subsystem the unitary matrix $V$ is just
$-i\sigma_{2}$. Thus, for arbitrary $j=0,1,2,3$ we have the
following relation:
\begin{equation}
V\sigma_{j}V^{\dagger}=\sigma_{2}\sigma_{j}\sigma_{2}=2\delta_{2j}\sigma_{j}-\sigma_{j}=(-1)^j
\sigma_{j},
\end{equation}
which, in turn, implies that $\varrho^{\tau_{A}}$ has the
following Hilbert-Schmidt representation:
\begin{equation}\label{HStauA}
\varrho^{\tau_{A}}=\frac{1}{2d}\left(\sum_{j=0}^{d^{2}-1} \xi_{0j}
\mathbbm{1}_{2} \ot f_j-\sum_{i=1}^{3}\sum_{j=0}^{d^{2}-1}
\xi_{ij}\sigma_i \ot f_j\right).
\end{equation}
Comparison of Eqs. (\ref{HS}) and (\ref{HStauA}) leads immediately
to the fact that
$\varrho+\varrho^{\tau_{A}}=\mathbbm{1}_{2}\ot\varrho_{B}$. Thus
the Beuer map (\ref{Bmap}) indeed gives zero when acting on the
two-dimensional subsystem. On the other hand, one recognizes in
this equality the equivalence between transposition and reduction
maps when both act on a $2\times 2$ matrix \cite{xor,reduction2},
i.e., $\Lambda_{r}(A)=\tau(A)$. Following Ref.
\cite{VollbrechtWolf} and using the above relations we may write
the following equalities for $\alpha\geq 1$:
\begin{eqnarray}\label{I.6}
\Tr\varrho_{B}^{\alpha}&=&\Tr\varrho(\mathbbm{1}_{2}\ot\varrho_{B}^{\alpha-1})=
\Tr\varrho(\mathbbm{1}_{2}\ot\varrho_{B})^{\alpha-1}\nonumber\\
&=&\Tr\varrho(\varrho+\varrho^{\tau_{A}})^{\alpha-1}.
\end{eqnarray}
Equation (\ref{I.6}), though seemingly not to be useful for
detecting entanglement, may be used to derive some inequalities
which are stronger than the entropic ones.

Before we make the general considerations for a natural
$\alpha\geq 2$ let us investigate the cases of $\alpha=3,\,4,\,5$
(for $\alpha=2$ the procedure presented beneath still holds,
however, it leads to the standard entropic inequality), since for
these values of $\alpha$ we do not need to make any assumptions.
We are going to show that omitting certain terms on the right-hand
side of Eq. (\ref{I.6}) one obtains inequalities stronger than the
respective entropic inequality.
For this purpose let us assume that $\varrho$ is separable, i.e.,
of the form (\ref{separable}).
Then the matrix
$\varrho^{\tau_{A}}=\sum_{i}p_{i}V\rho_{i}^{T}V^{\dagger}\ot
\tilde{\rho}_{i}$ is obviously positive since
$V\rho_{i}^{T}V^{\dagger}\geq 0$ for $\rho_{i}\geq 0$. Moreover,
let us recall the fact that even though the product of two
positive matrices $A\geq 0$ and $B\geq 0$ need not be a positive
matrix, the trace of the product is always nonnegative, i.e., $\Tr
AB\geq 0$ \cite{footnote1}.
In further considerations we also apply the fact that, in general, terms such as
\begin{equation}\label{terms}
Q_{l_{1},\ldots,l_{n}}^{k_{1},\ldots,k_{n}}(\varrho)=
\Tr\left[\varrho^{l_{1}}(\varrho^{\tau_{A}})^{k_{1}}\varrho^{l_{2}}(\varrho^{\tau_{A}})^{k_{2}}\ldots
\varrho^{l_{n}}(\varrho^{\tau_{A}})^{k_{n}}\right]
\end{equation}
with $l_{i},k_{i}\in\mathbb{N}$ and odd $k_{1}+\ldots+k_{n}$ are
negative for some entangled states. The negativity of terms such
as Eq. (\ref{terms}) may be easily seen in case of a
$d$-dimensional maximally entangled state
\begin{equation}
\varrho=P_{+}^{(d)}=\frac{1}{d}
\sum_{i,j=0}^{d-1}\ket{ii}\bra{jj}.
\end{equation}
First, one sees that $\varrho^{l_{j}}=\varrho$ and
$P_{+}^{(d)T_{A}}=(1/d)\mathscr{V}^{(2)}$, where
$\mathscr{V}^{(2)}$ is the known swap operator defined as
$\mathscr{V}^{(2)}\ket{\Phi_{1}}\ket{\Phi_{2}}=\ket{\Phi_{2}}\ket{\Phi_{1}}$
with $\ket{\Phi_{1(2)}}\in\mathbb{C}^{d}$. Secondly, the
Hermiticity and unitarity of $\mathscr{V}^{(2)}$ allow us to write
that $(\varrho^{\tau_{A}})^{k_{j}}=(1/d^{k_{j}})\mathbbm{1}_{d}$
whenever $k_{j}$ is even, and
$(\varrho^{\tau_{A}})^{k_{j}}=(1/d^{k_{j}})
(V\ot\mathbbm{1}_{d})\mathscr{V}^{(2)}(V^{\dagger}\ot\mathbbm{1}_{d})=(1/d^{k_{j}-1})P_{+}^{(d)\tau_{A}}$
for odd $k_{j}$. Therefore the expression (\ref{terms}) reduces to
\begin{equation}\label{terms2}
Q_{l_{1},\ldots,l_{n}}^{k_{1},\ldots,k_{n}}(P_{+}^{(d)})
=\frac{\Tr
(P^{(d)}_{+}P_{+}^{(d)\tau_{A}})^{r}}{d^{k_{1}+\ldots+k_{n}-r}}=\frac{\Tr
(P^{(d)}_{+}P_{+}^{(d)\tau_{A}}P^{(d)}_{+})^{r}}{d^{k_{1}+\ldots+k_{n}-r}},
\end{equation}
where $r$ is an odd number. Moreover,
$P^{(d)}_{+}P_{+}^{(d)\tau_{A}}P_{+}^{(d)}=-(1/d)P^{(d)}_{+}$, which
makes the expression in Eq. (\ref{terms2}) negative.

Now let us consider the special cases of Eq. (\ref{I.6}). For
$\alpha=3$ we obtain
\begin{equation}\label{I.8}
\Tr\varrho_{B}^{3}=\Tr\varrho^{3}+2\Tr\varrho^{2}\varrho^{\tau_{A}}+\Tr\varrho(\varrho^{\tau_{A}})^{2}.
\end{equation}
Since for any natural $n$ and separable state $\varrho$ the
matrices $\varrho^{\tau_{A}}$ and $\varrho^{n}$ are positive, one
concludes that $\Tr\varrho^{n}\varrho^{\tau_{A}}\geq 0$. Thus,
under the assumption that $\varrho$ is separable one may omit the
term $\Tr\varrho^{2}\varrho^{\tau_{A}}$, obtaining the following
inequality:
\begin{equation}
\Tr\varrho_{B}^{3}\geq \Tr\varrho^{3}+\Tr\varrho(\varrho^{\tau_{A}})^{2}.
\end{equation}
Since $\Tr\varrho(\varrho^{\tau_{A}})^{2}\geq 0$ even for
entangled states, one could see that this inequality is more
powerful than its entropic counterpart $\Tr\varrho_{B}^{3}\geq
\Tr\varrho^{3}$.

In an analogous way one may derive an inequality for $\alpha=4$.
From Eq. (\ref{I.6}) one has
\begin{eqnarray}
&&\hspace{-0.7cm}\Tr\varrho_{B}^{4}=\Tr\varrho^{4}+3\Tr\varrho^{3}\varrho^{\tau_{A}}+2\Tr\varrho^{2}(\varrho^{\tau_{A}})^{2}\nonumber\\
&&\hspace{0.5cm}+\Tr(\varrho\varrho^{\tau_{A}})^{2}+\Tr\varrho(\varrho^{\tau_{A}})^{3}.
\end{eqnarray}
The term $\Tr(\varrho\varrho^{\tau_{A}})^{2}$ is always positive
since
$\Tr(\varrho\varrho^{\tau_{A}})^{2}=\Tr(\sqrt{\varrho}\varrho^{\tau_{A}}\sqrt{\varrho})^{2}$.
Now, omitting the terms with odd number of $\varrho^{\tau_{A}}$ in
the product, which may be negative for some entangled states, one
obtains
\begin{equation}
\Tr\varrho_{B}^{4}\geq
\Tr\varrho^{4}+2\Tr\varrho^{2}(\varrho^{\tau_{A}})^{2}
+\Tr(\varrho\varrho^{\tau_{A}})^{2}.
\end{equation}
Again, this inequality must be stronger than its entropic
counterpart since all terms in the above are positive.

Finally, for $\alpha=5$ from Eq. (\ref{I.6}) we obtain
\begin{eqnarray}
\Tr\varrho_{B}^{5}&=&\Tr\varrho^{5}+4\Tr\varrho^{4}\varrho^{\tau_{A}}+
3\Tr\varrho^{3}(\varrho^{\tau_{A}})^{2}+3\Tr\varrho^{2}\varrho^{\tau_{A}}\varrho\varrho^{\tau_{A}}\nonumber\\
&&+2\Tr\varrho^{2}(\varrho^{\tau_{A}})^{3}+2\Tr\varrho\varrho^{\tau_{A}}
\varrho(\varrho^{\tau_{A}})^{2}
+\Tr\varrho(\varrho^{\tau_{A}})^{4}.\nonumber\\
\end{eqnarray}
For separable $\varrho$ terms in which $\varrho^{\tau_{A}}$ occurs
in odd powers may be omitted since they are positive. The terms
$\Tr\varrho^{4}\varrho^{\tau_{A}}$ and
$\Tr\varrho^{2}(\varrho^{\tau_{A}})^{3}$ are positive for
separable states. Moreover, we have
\begin{eqnarray}\label{nier}
\Tr\varrho\varrho^{\tau_{A}}\varrho(\varrho^{\tau_{A}})^{2}&=&
\Tr\varrho^{\tau_{A}}\varrho\varrho^{\tau_{A}}\varrho^{\tau_{A}}\varrho\nonumber\\
&=&\Tr\varrho^{\tau_{A}}(\varrho^{\tau_{A}}\varrho)^{\dagger}(\varrho^{\tau_{A}}\varrho)\geq 0.
\end{eqnarray}
Now, omitting the mentioned terms, we have
\begin{eqnarray}\label{nier5}
\Tr\varrho^{5}_{B}&\geq& \Tr\varrho^{5}+3\Tr\varrho^{3}(\varrho^{\tau_{A}})^{2}
+3\Tr\varrho^{2}\varrho^{\tau_{A}}\varrho\varrho^{\tau_{A}}\nonumber\\
&&+\Tr\varrho(\varrho^{\tau_{A}})^{4}.
\end{eqnarray}
Since
\begin{eqnarray}
\Tr\varrho^{2}\varrho^{\tau_{A}}\varrho\varrho^{\tau_{A}}&=&\Tr\varrho\varrho^{\tau_{A}}\varrho\varrho^{\tau_{A}}\varrho\nonumber\\
&=& \Tr(\sqrt{\varrho}\varrho^{\tau_{A}}\varrho)^{\dagger}(\sqrt{\varrho}\varrho^{\tau_{A}}\varrho)\geq 0,
\end{eqnarray}
all the terms appearing in the inequality (\ref{nier5}) are positive even for entangled states and
thus again the inequality (\ref{nier5}) is stronger than the respective entropic one.

It should be clarified that our aim is to leave on the right-hand
side of the derived inequalities only the terms that remain
positive, even if partial time reversal of a state is not a
positive matrix. Then the possibility of violation of the
respective inequalities by entangled states is stronger. In
general (i.e., for natural $\alpha\geq 6$) it is not clear which
terms of the form (\ref{terms}) are positive when
$\varrho^{\tau_A}$ is positive and which could become negative for
NPT states. Therefore, in general, we do not know which terms can
be removed on the right-hand side of Eq. (\ref{I.6}) to obtain the
strong inequalities for higher $\alpha$. Hence to derive the
inequalities for arbitrary $\alpha\in\mathbb{N}\setminus\{0\}$ we
make an additional assumption that
$[\varrho,\varrho^{\tau_{A}}]=0$, which
by virtue of the fact that
$[\mathbbm{1}_{2}\ot\varrho_{B},\varrho]^{\tau_{A}}=[\mathbbm{1}_{2}\ot\varrho_{B},\varrho^{\tau_{A}}]$,
is equivalent to the condition
$[\mathbbm{1}_{2}\ot\varrho_{B},\varrho]=0$. Now we state the
general criterion for states acting on
$\mathbb{C}^{2}\ot\mathbb{C}^{d}$ as the following fact.

{\it Fact 1.-} Let $\varrho$ represent a separable state defined
on $\mathbbm{C}^2\otimes\mathbbm{C}^d$ commuting with
$\mathbbm{1}_{2}\ot\varrho_{B}$. Then for
$\alpha\in\mathbbm{N},\,\alpha\geq 1$ the following inequality
holds
\begin{equation}\label{kunity}
\Tr\varrho_{B}^{\alpha}\geq \sum_{k=0}^{\lfloor
(\alpha-1)/2\rfloor} \left(\begin{array}{c}
\alpha-1\\
2k
\end{array}\right)\Tr\varrho^{\alpha-2k}(\varrho^{\tau_{A}})^{2k}.
\end{equation}
The proof of the fact is rather straightforward and follows from
the commutativity of $\varrho$ and $\varrho^{\tau_{A}}$ and the
known Newton binomial formula.

For the sake of simplicity the above may be rewritten as
\begin{equation}\label{2N1}
\Tr\varrho_{B}^{\alpha}\geq\frac{1}{2}
\left[\Tr{\varrho(\varrho+\varrho^{\tau_A})^{\alpha-1}}+\Tr{\varrho(\varrho-\varrho^{\tau_A})^{\alpha-1}}\right].
\end{equation}

On the other hand, to show that this inequality is stronger than
the entropic one, it may also be rewritten as follows:
\begin{equation}\label{Fact1}
\Tr\varrho_{B}^{\alpha}\geq
\Tr\varrho^{\alpha}+\sum_{k=1}^{\lfloor (\alpha-1)/2\rfloor}
\left(\begin{array}{c}
\alpha-1\\
2k
\end{array}\right)\Tr\varrho^{\alpha-2k}(\varrho^{\tau_{A}})^{2k}.
\end{equation}
Since the second term in the above is always positive (even for
entangled $\varrho$), all the inequalities for arbitrary natural
$\alpha>2$ are stronger than their entropic counterparts (note,
that for $\alpha=2$ the above inequality becomes the standard
entropic inequality).

{\it Remark 1.1.} One should note that if $\varrho_{A}$ is
non-degenerate and $\varrho_{A} \ot \mathbbm{1}_{d}$ commutes with
the state $\varrho$, then immediately $\varrho$ must be separable.
This follows form the fact that then the latter has to have all
its eigenvectors of the separable form $\ket{\phi} \ot \ket{\Psi}$
where $\ket{\phi}$ is an eigenvector of $\varrho_{A}$ and
$\ket{\Psi}$ is some vector from the Hilbert space describing the
second subsystem.

{\it Remark 1.2.} One could easily see that in case $d=2$, i.e.,
two-qubit states its is possible to derive a dual inequality of
Eq. (\ref{Fact1}) with map $\tau$ acting on the subsystem $B$,
which is
\begin{equation}\label{2N2}
\Tr\varrho_{A}^{\alpha}\geq\frac{1}{2}
\left[\Tr\varrho(\varrho+\varrho^{\tau_{B}})^{\alpha-1}+\Tr\varrho(\varrho-\varrho^{\tau_{B}})^{\alpha-1}\right].
\end{equation}

{\it Remark 1.3.} It should also be emphasized that Eq.
(\ref{2N1}) leads to a stronger inequality, from which, however,
it seems impossible to construct the many copy entanglement
witnesses. The inequality follows from the observation that for
separable states $\varrho^{\tau_{A(B)}}\geq 0$, which implies that
$\varrho^{\tau_{A(B)}}=|\varrho^{\tau_{A(B)}}|$. Hence one may
rewrite Eq. (\ref{2N1}) as
\begin{equation}
\Tr\varrho_{B}^{\alpha}\geq\frac{1}{2}\left[\Tr\varrho(\varrho+\left|\varrho^{\tau_{A}}\right|)^{\alpha-1}
+\Tr\varrho(\varrho-\varrho^{\tau_{A}})^{\alpha-1}\right].
\end{equation}
We add the absolute value only in the first term since it can
increase the right-hand side, while in case of the second term the
addition of the absolute value could decrease it.

\begin{figure}
(a)\includegraphics[width=0.23\textwidth]{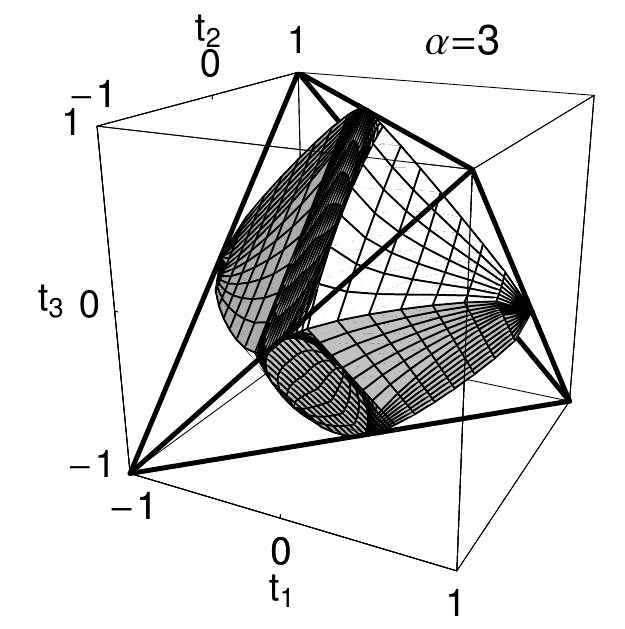}\includegraphics[width=0.23\textwidth]{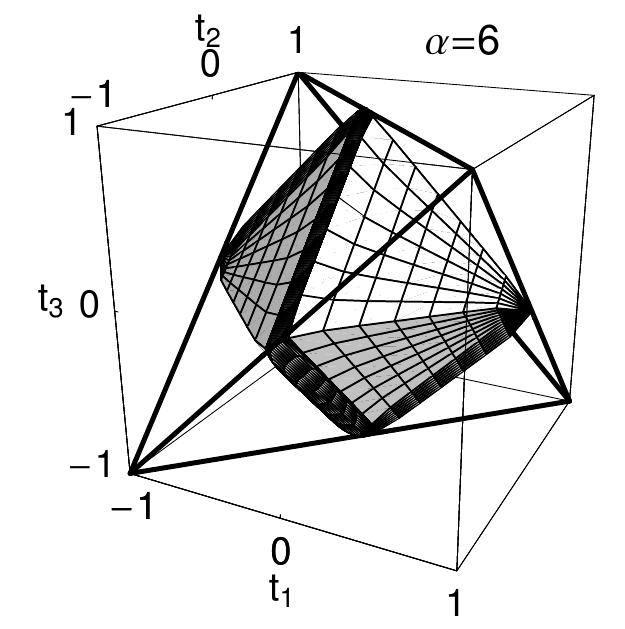}\\
(b)\includegraphics[width=0.23\textwidth]{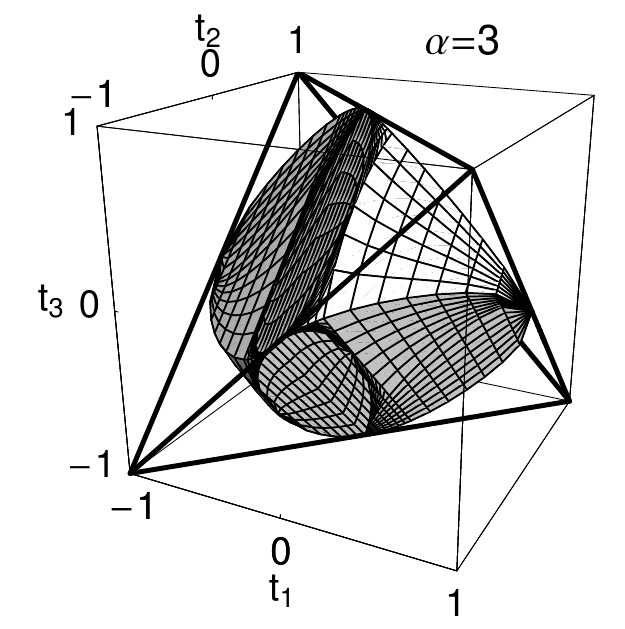}\includegraphics[width=0.23\textwidth]{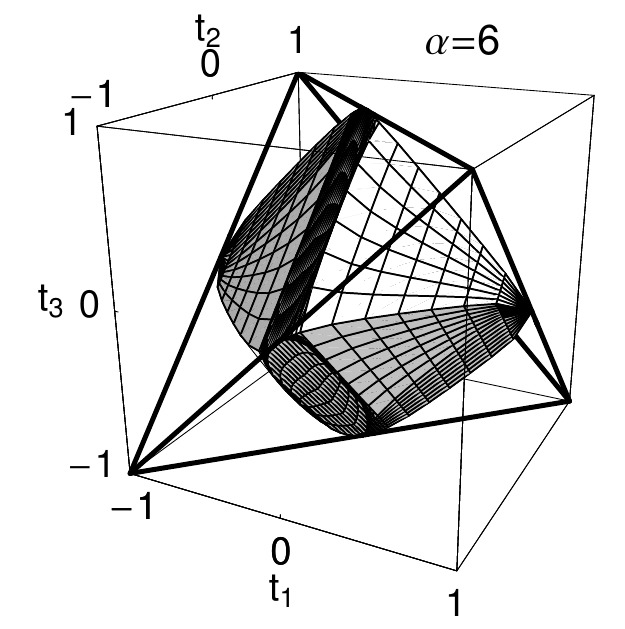}\\
(c) \includegraphics[width=0.23\textwidth]{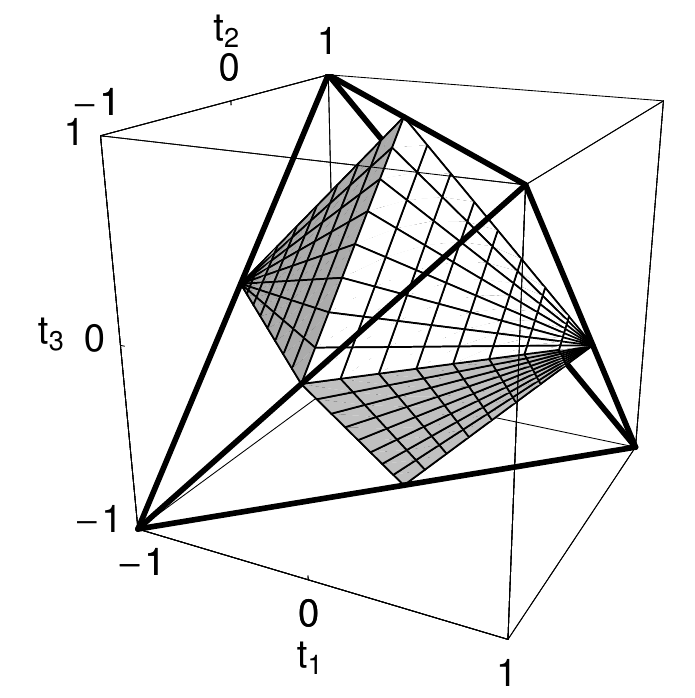}
\caption{Comparison of inequality (\ref{kunity}) and that proposed
in Ref. \cite{Guhne} in the case of Bell-diagonal states. In both
upper figures (a) the region which satisfies our inequalities is
presented (left for $\alpha=3$ and right for $\alpha=6$), while in
figures (b) the states satisfying inequalities from Ref.
\cite{Guhne} are shown (for the same values of $\alpha$). For
comparison in figure (c) the tetrahedron of all the Bell-diagonal
states and octahedron containing Bell-diagonal separable states
are displayed \cite{MHRH}.}\label{belldiag}
\end{figure}

To show the effectiveness of Eq. (\ref{kunity}) we consider two
classes of two-qubit states. The first are the two-qubit
Bell-diagonal states
\begin{equation}
\varrho_{\mathrm{Bell}}(p,q,r)=pP_{+}+qP_{-}+rQ_{+}+(1-p-q-r)Q_{-},
\end{equation}
where $P_{\pm}$ and $Q_{\pm}$ are projectors onto Bell states
$\ket{\psi_{\pm}}=(1/\sqrt{2})(\ket{01}\pm\ket{10})$ and $\ket{\phi_{\pm}}=(1/\sqrt{2})(\ket{00}\pm\ket{11})$, respectively. Bell-diagonal states have a simple form \cite{MHRH} in terms of the Pauli matrices (\ref{Pauli}):
\begin{equation}
\varrho_{\mathrm{Bell}}(\mathbf{t})=\frac{1}{16}
\left(\mathbbm{1}_{2}\ot\mathbbm{1}_{2}+\sum_{i=1}^{3}t_{i}\sigma_{i}\ot\sigma_{i}\right),
\end{equation}
where $t_{i}\in\mathcal{V}$ $(i=1,2,3)$ and $\mathcal{V}$ is a
tetrahedron with vertices $(-1,-1,-1)$, $(-1,1,1)$, $(1,-1,1)$,
and $(1,1,-1)$ corresponding to all four two-qubit Bell states
(see Fig. \ref{belldiag}c).

We compare Eq. (\ref{kunity}) to the one derived in Ref.
\cite{Guhne}, i.e.,
\begin{equation}\label{Lew}
S^{T}_{\alpha}(M)_{\varrho}\geq \frac{1-2^{1-\alpha}}{\alpha-1},
\end{equation}
where $M$ denotes the Bell-diagonal observable with non-degenerate
spectrum and $S_{\alpha}^T$ stands for the Tsallis entropy of the
classical probability distribution. The Tsallis entropy of a
probability distribution $\mathcal{P}=(p_1,\ldots,p_n)$ is defined
as $S_{\alpha}^{T}=[1-\sum_k (p_k)^{\alpha}]/(\alpha - 1)$.
The results obtained for $\alpha=3$ and $\alpha=6$ (Fig.
\ref{belldiag}) show that the region of states not detected by Eq.
(\ref{kunity}) is smaller than the one derived from inequality
(\ref{Lew}).

\begin{figure}[!hbp]
\includegraphics[width=0.23\textwidth]{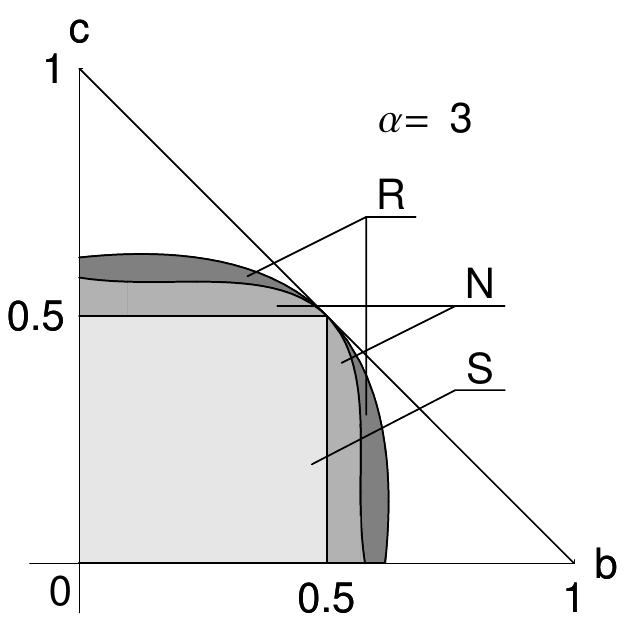}\includegraphics[width=0.23\textwidth]{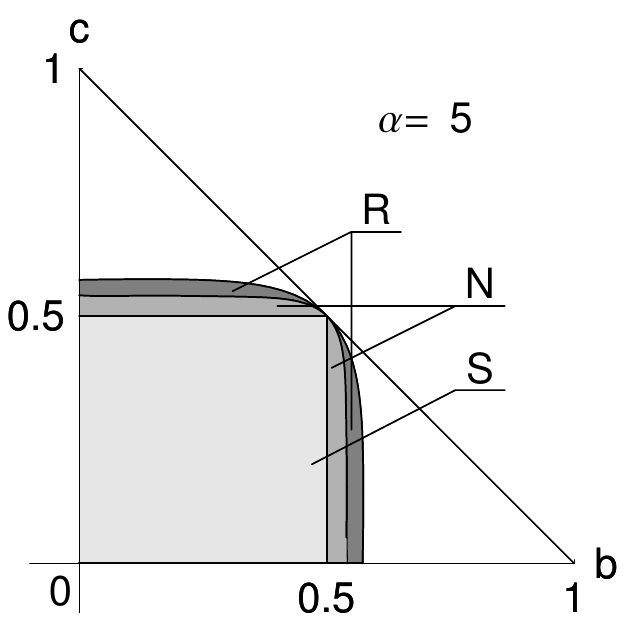}\\
\caption{The comparison of entropic inequalities and the present
ones in the case of states given by (\ref{NPT}). The triangle
specifies the range of parameters $b$ and $c$ for which Eq.
(\ref{NPT}) represents a state. Square $S$ denotes a subset of
separable states. States which are not detected by standard
entropic inequalities are represented by darker gray set marked
with $R$, while the brighter gray set marked with $N$ indicates
the states which are not detected by the inequality
(\ref{kunity}). The sets overlap in the region near separable
states (i.e., $S\subset N\subset R$). From the analysis one
concludes that the set of entangled states detected by our
inequalities is considerably bigger than the set corresponding to
entropic inequalities. Plots are made for $\alpha=3$ and
$\alpha=5$.}\label{divqubit}
\end{figure}

The second class are the two-parameter states considered in Ref.
\cite{DiV}:
\begin{equation}\label{NPT}
\tilde{\varrho}(b,c)=a\sum_{i=0}^{1}\proj{ii}+b\proj{\psi_{-}}+c\proj{\psi_{+}},
\end{equation}
where $\ket{\psi_{\pm}}$ are defined as previously and
$a=(1/2)(1-b-c)$. One can easily check that
$\Tr_{B}\tilde{\varrho}(b,c)=\Tr_{A}\tilde{\varrho}(b,c)=(1/2)\mathbbm{1}_{2}$ and the assumption of
Fact 1 is satisfied. Comparison
with the entropic inequalities for $\alpha=3$ and $\alpha=5$ is shown
in Fig. \ref{divqubit}.

\subsection{General scalar inequalities.}\label{sub-2.B}
In the paragraph we generalize the above results to bipartite
systems with arbitrarily dimensional subsystems. The property
$\mathbbm{1}_{2}\ot \varrho_{B}=\varrho+\varrho^{\tau_{A}}$
possessed by states on the Hilbert space
$\mathbb{C}^{2}\ot\mathbb{C}^{d}$ is in general not valid for
systems defined on $\mathbb{C}^{d_{A}}\ot\mathbb{C}^{d_{B}}$.
However, the separability criterion based on the general map
$\Lambda_{U}^{(-)}$ provides us with the operator inequalities
$\varrho_{A}\ot \mathbbm{1}_{d_{B}}\geq
\varrho+\varrho^{\tau_{B}^{U}}$ and
$\mathbbm{1}_{d_{A}}\ot\varrho_{B}\geq
\varrho+\varrho^{\tau_{A}^{U}}$, which are true for an arbitrary
bipartite separable state. In the following fact we propose an
inequality resulting from the Breuer-Hall map.

{\it Fact 2.} If a given state $\varrho$ on
$\mathbb{C}^{d_{A}}\otimes\mathbb{C}^{d_{B}}$ is separable and has
the property that $[\varrho,\varrho_{A}\ot\mathbbm{1}_{d_{B}}]=0$
$([\varrho,\mathbbm{1}_{d_{A}}\ot\varrho_{B}]=0)$ then for an
arbitrary natural number $\alpha \geq  1$ \cite{footnote2}
\begin{equation}\label{1tr}
\Tr\varrho_{A(B)}^{\alpha}\geq
\Tr\varrho\left(\varrho+\varrho^{\tau_{B(A)}^{U}}\right)^{\alpha-1}.
\end{equation}

{\it Proof.} The proof is a simple consequence of few well known
facts. Let $\varrho$ be a state obeying the assumptions of the
theorem. Then, since in general
\begin{equation}
[\varrho,\varrho_{A}\ot\mathbbm{1}_{d_{B}}]^{\tau_{B}^{U}}=
[\varrho^{\tau_{B}^{U}},\varrho_{A}\ot\mathbbm{1}_{d_{B}}],
\end{equation}
the assumption $[\varrho,\varrho_{A}\ot\mathbbm{1}_{d_{B}}]=0$
implies that also
$[\varrho^{\tau_{B}^{U}},\varrho_{A}\ot\mathbbm{1}_{d_{B}}]=0$.
Therefore one has
\begin{equation}\label{nier_op}
\varrho_{A}^{\alpha-1}\ot\mathbbm{1}_{d_{B}}=(\varrho_{A}\ot\mathbbm{1}_{d_{B}})^{\alpha-1}\geq
\left(\varrho+\varrho^{\tau_{B}^{U}}\right)^{\alpha-1}.
\end{equation}
Exploiting the property that if $A\geq B$ then also
$X^{\dagger}A X\geq X^{\dagger} B X$ for an arbitrary matrix $X$ we may write
\begin{equation}
\sqrt{\varrho}(\varrho_{A}\ot\mathbbm{1}_{d_{B}})^{\alpha-1}\sqrt{\varrho}\geq
\sqrt{\varrho}\left(\varrho+\varrho^{\tau_{B}^{U}}\right)^{\alpha-1}\sqrt{\varrho}.
\end{equation}
Finally, since if $A\geq B$ then also $\Tr A\geq \Tr B$, we obtain the postulated inequality
\begin{equation}
\Tr\varrho(\varrho_{A}\ot\mathbbm{1}_{d_{B}})^{\alpha-1}=\Tr\varrho_{A}^{\alpha}\geq
\Tr\varrho\left(\varrho+\varrho^{\tau_{B}^{U}}\right)^{\alpha-1},
\end{equation}
finishing the proof. $\square$

{\it Remark 2.1.} Assuming that
$[\varrho_{A}\ot\mathbbm{1}_{d_{B}},\varrho]=0$ the operator
inequality in Eq. (\ref{nier_op}) itself may lead to a criterion
detecting entanglement, which is stronger than the one for
$\alpha=2$, i.e., following from the linear map. Analysis of these
inequalities will be made in one of the next subsections.

{\it Remark 2.2.} Assuming again that $\varrho$ represents a
separable state one has
$\big|\varrho^{\tau_{A(B)}^{U}}\big|=\varrho^{\tau_{A(B)}^{U}}$
and therefore
\begin{equation}\label{module}
\Tr\varrho_{A(B)}^{\alpha}\geq
\Tr\varrho\left(\varrho+\left|\varrho^{\tau_{B(A)}^{U}}\right|\right)^{\alpha-1}.
\end{equation}
This inequality is stronger than the previous one, however, since
it contains an absolute value of a matrix, it is, to our
knowledge, not measurable on few copies of state.

The generalization of Fact 1 to higher-dimensional states is also
possible. Let us state it as the following fact.

{\it Fact 3.} Assume that $\varrho$ is a separable state acting on
$\mathbb{C}^{d_{A}}\otimes\mathbb{C}^{d_{B}}$ and that the
commutator $\big[\varrho,\mathbbm{1}_{d_{A}}\ot\varrho_{B}\big]$
disappears, then for a natural number $\alpha\geq 1$
\begin{equation}\label{symmetric}
\Tr\varrho_{A}^{\alpha}\geq
\frac{1}{2}\left[\Tr\varrho\left(\varrho+\varrho^{\tau_B^{U}}\right)^{\alpha-1}
+\Tr\varrho\left(\varrho-\varrho^{\tau_B^{U}}\right)^{\alpha-1}\right].
\end{equation}
{\it Proof.} Let us consider the map $\Lambda_{U}^{(+)}$
introduced at the beginning of the section, Eq. (\ref{phimap}). It
leads to the separability criterion
$\mathbbm{1}_{d_{A}}\ot\varrho_{B}\geq\varrho-\varrho^{\tau_{A}^U}$.
Now, applying the methods used in the proof of Fact 2 to the
criterion resulting from $\Lambda_{U}^{(+)}$ we obtain the
following inequality
\begin{equation}\label{phiineq}
\Tr\varrho_{B}^{\alpha}\geq
\Tr\varrho\left(\varrho-\varrho^{\tau_{A}^{U}}\right)^{\alpha-1},
\end{equation}
which is fulfilled by all separable states satisfying the
commutativity assumption. Combining the inequalities (\ref{1tr})
and (\ref{phiineq}) we obtain the inequality (\ref{symmetric}).
Note that the analogous inequality can be also derived for the
second subsystem.$\blacksquare$

{\it Remark 3.1} For states $\varrho$ that commute with
$\varrho^{\tau_{A}^{U}}$ (i.e.
$\big[\varrho,\varrho^{\tau_{A}^{U}}\big]=0$) and for natural
$\alpha\geq 2$ some terms on the right-hand side of the inequality
(\ref{1tr}) can be removed, leading to general inequalities of the
form
\begin{equation}\label{4tr}
\Tr\varrho_{A(B)}^{\alpha}\geq\Tr\varrho^{\alpha}+G_{\alpha}(\varrho),
\end{equation}
where $G_{\alpha}(\varrho)$ is a sum of terms of type
(\ref{terms}) such that the inequality remains true for all
separable states. Note that the same procedure was proposed in the
previous subsection.
Since again, the above represents somehow improved entropic
inequality it should, in principle, be more powerful, whenever
$G_{\alpha}(\varrho)$ is positive for any entangled state
$\varrho$.
Note that the inequality proposed in Fact 3 is also of this form
%

It is interesting to analyze the limit $\alpha \to \infty$ for the
inequality (\ref{symmetric}). It can be easily done since one
assumes that $\big[\varrho,\varrho^{\tau_A^U}\big]=0$. Let us
transform Eq. (\ref{symmetric}) to the following form
\begin{equation}\label{sym_log}
\frac{\log\Tr\varrho_{B}^{\alpha}}{1-\alpha}\leq\frac{\log
\displaystyle\left\{\frac{1}{2}\Tr\varrho\left[\left(\varrho+\varrho^{\tau_A^{U}}\right)^{\alpha-1}+
\left(\varrho-\varrho^{\tau_A^{U}}\right)^{\alpha-1}\right]\right\}}{1-\alpha}.
\end{equation}
The left-hand side of the above inequality is the Renyi entropy of
subsystem $\varrho_B$, and due to Ref. \cite{MHPH} in the limit
$\alpha \to \infty$ gives $-\log ||\varrho_B||$, where
$||\varrho_B||$ is an operator norm of $\varrho_B$.
Due to the assumption that
$\big[\varrho,\varrho^{\tau_{A}^{U}}\big]=0$, there exists a
common orthonormal basis of eigenvectors of $\varrho$ and
$\varrho^{\tau_A^U}$. We denote it by $\{\ket{\psi_i}\}$. Then
$\varrho+\varrho^{\tau_A^U}$ and $\varrho-\varrho^{\tau_A^U}$ must
have the same eigenvectors as $\varrho$.

Let $\lambda_i$, $\lambda_i^{\tau}$, $\lambda^{+}_i$,
$\lambda^{-}_i$ denote the eigenvalues of $\varrho$,
$\varrho^{\tau_A^U}$, $\varrho+\varrho^{\tau_A^U}$ and
$\varrho-\varrho^{\tau_A^U}$, corresponding to eigenvector
$\ket{\psi_i}$. We can than rewrite Eq. (\ref{sym_log}) as
\begin{eqnarray}\label{41}
\frac{\log\Tr\varrho_{B}^{\alpha}}{1-\alpha}
\leq\frac{\log\left\{\displaystyle\frac{1}{2}\displaystyle\sum_{i}\lambda_i
\left[(\lambda^{+}_i)^{\alpha-1}+(\lambda^{-}_i)^{\alpha-1}\right]\right\}}
{1-\alpha}.
\end{eqnarray}
We need to show that the argument of logarithm is positive.
Henceforward we will assume that all $\lambda_{i}$ are strictly
positive since terms with $\lambda_i=0$ do not contribute to the
sum under logarithm. Therefore one sees that
\begin{equation}
\sum_{i}\lambda_i
\left[(\lambda^{+}_i)^{\alpha-1}+(\lambda^{-}_i)^{\alpha-1}\right]\geq0
\end{equation}
and since all terms in the sum are nonnegative the equality is
possible only if
$(\lambda^{+}_i)^{\alpha-1}+(\lambda^{-}_i)^{\alpha-1}=0$ for all
$i$. This, however, is impossible since all such terms are of the
form
$(\lambda_i+\lambda_i^{\tau})^{k}+(\lambda_i-\lambda_i^{\tau})^{k}$
with $\lambda_i>0$ and $\lambda_i^{\tau}\geq 0$. Now the
positivity can be seen by a straightforward calculation using the
binomial formula.

We introduce the following notation
$\lambda^{+}_{\mathrm{max}}=\max_i \{|\lambda_i^{+}|\}$,
$\lambda^{-}_{\mathrm{max}}=\max_i \{|\lambda_i^{-}|\}$,
remembering that we exclude the situation $\lambda_i=0$. So
$\lambda^{\pm}_{\mathrm{max}}$ are the maximum eigenvalues of
$|\varrho\pm\varrho^{\tau_A^U}|$ corresponding to a nonzero
$\lambda_i$. By $q_{\max}$ we denote
$\max\{\lambda^{+}_{\mathrm{max}},\lambda^{-}_{\mathrm{max}}\}$.
Moreover, let $\tilde{\lambda}^{+}_{i}=\lambda^{+}_{i}/q_{\max}$
and $\tilde{\lambda}^{-}_{i}=\lambda^{-}_{i}/q_{\max}$. Now we may
rewrite Eq. (\ref{41}) as
\begin{equation}
\frac{\log\Tr\varrho_{B}^{\alpha}}{1-\alpha}\leq \frac{\log
q_{\mathrm{max}}^{\alpha-1}}{1-\alpha}+\frac{\log\frac{1}{2}\left\{\displaystyle\sum_{i}\lambda_i
\left[(\tilde{\lambda}^{+}_{i})^{\alpha-1}+(\tilde{\lambda}^{-}_{i})^{\alpha-1}\right]\right\}}
{1-\alpha}
\end{equation}
and finally as
\begin{equation}
\frac{\log\Tr\varrho_{B}^{\alpha}}{1-\alpha}\leq-\log
q_{\mathrm{max}}
+\frac{\log\frac{1}{2}\left\{\displaystyle\sum_{i}\lambda_i
\left[(\tilde{\lambda}_{i}^{+})^{\alpha-1}+(\tilde{\lambda}_{i}^{-})^{\alpha-1}\right]\right\}}
{1-\alpha}.
\end{equation}
It should be mentioned that the logarithm in the second term on
right-hand side of the above inequality is finite in the limit
$\alpha\to\infty$ since $\sum_{i}\lambda_i
\big[(\tilde{\lambda}_{i}^{+})^{\alpha-1}+(\tilde{\lambda}_{i}^{-})^{\alpha-1}\big]$
is bounded from above and can never approach zero when
$\alpha\to\infty$.

Zero under logarithm can only come from a term such as
$1^{\alpha}+(-1)^{\alpha}$ which is equivalent to
$\lambda_{\mathrm{max}}^{+}=\lambda_{\mathrm{max}}^{-}$ and
$\lambda_i^{-}=-\lambda_{\mathrm{max}}^{-}$ for some $i$ (let us
denote this particular index by $N$). This, in turn, could happen
only if
$\lambda_{N}+\lambda_{N}^{\tau}=-(\lambda_{N}-\lambda_{N}^{\tau})$
leading to $\lambda_{N}=0$. Such situation, however, was excluded
at the outset. Thus, one sees that the logarithm is always finite
and taking the limit $\alpha\to\infty$ on both sides we obtain
\begin{equation}
-\log{||\varrho_B||}\leq-\log q_{\mathrm{max}},
\end{equation}
which can be also written as
\begin{equation}
||\varrho_B||\geq
\max\{\lambda_\mathrm{max}^{+},\lambda_\mathrm{max}^{-}\}.
\end{equation}
A similar inequality one may derive for $\varrho_{A}$. Moreover,
comparison with Eq. (\ref{entropic_niek}) shows that the just
derived inequality must be stronger than its entropic counterpart.
Let us now present the second inequality which is also based on
Breuer-Hall map, however, its derivation is a little bit more
involving.

{\it Fact 4.} Assume that $\varrho$ acting on
$\mathbb{C}^{d_{A}}\ot\mathbb{C}^{d_{B}}$ is separable and
$[\varrho,\varrho^{\tau_{A(B)}^{U}}]=0$ with a given antisymmetric
unitary $U$. Then for $\alpha>1$ the following inequality holds:
\begin{equation}\label{iloczyn}
\Tr\varrho_{B(A)}^{\alpha}\geq
2^{\alpha-1}\left[\Tr\varrho^{\frac{\alpha+1}{2}}\left(\varrho^{\tau_{A(B)}^{U}}\right)^{\frac{\alpha-1}{2}}\right].
\end{equation}
{\it Proof.} The proof goes along the same lines as presented in
Ref. \cite{VollbrechtWolf}. First we may write
\begin{equation}
\Tr\varrho_{B}^{\alpha}=\Tr\varrho(\mathbbm{1}_{d_{A}}\ot\varrho_{B})^{\alpha-1}=\Tr
e^{\log\varrho}e^{(\alpha-1)\log\mathbbm{1}_{d_{A}}\ot\varrho_{B}}.
\end{equation}
Now, since $\Tr\, e^{A}e^{B}\geq \Tr\, e^{A+B}$ (see Ref.
\cite{Bhatia}) and due to the equation
$\mathbbm{1}_{d_{A}}\ot\varrho_{B}\geq\varrho+\varrho^{\tau_{A}^{U}}$
and monotonicity of the logarithm, we have
\begin{equation}
\Tr\varrho_{B}^{\alpha}\geq\Tr
e^{\log\varrho+(\alpha-1)\log\big(\varrho+\varrho^{\tau_{A}^{U}}\big)}.
\end{equation}
Then we may use concavity of the logarithm to obtain
\begin{equation}\label{ineq}
\Tr\varrho_{B}^{\alpha}\geq 2^{\alpha-1}\Tr\,e^{[(\alpha+1)/2]
\log\varrho+[(\alpha-1)/2]\log\varrho^{\tau_{A}^{U}}}.
\end{equation}
Finally by virtue of the assumption that $\varrho$ and
$\varrho^{\tau_{A}^{U}}$ commute we have the commutativity of
their logarithms, and therefore
\begin{equation}\label{Fact3}
\Tr\varrho^{\alpha}_{B}\geq
2^{\alpha-1}\Tr\left[\varrho^{\frac{\alpha+1}{2}}\left(\varrho^{\tau_{A}^{U}}\right)^{\frac{\alpha-1}{2}}\right],
\end{equation}
finishing the proof.

{\it Remark 4.1.} The remark here is that in the above inequality
for $\alpha=2k$ one gets the square roots of
$\varrho^{\tau_{A}^{U}}$, which in case of entangled states may
lead to complex eigenvalues. Moreover, the inequality may be
strengthened by taking only the even powers of
$\varrho^{\tau_{A}^{U}}$, since in such case the RHS would remain
positive even for entangled states. Therefore we assume that
$\alpha=4k+1$. Then the inequality may be rewritten as
\begin{equation}\label{48}
\Tr\varrho^{4k+1}_{B(A)}\geq
2^{4k}\Tr\left[\varrho^{2k+1}\left(\varrho^{\tau_{A(B)}^{U}}\right)^{2k}\right] \quad (k=0,1,\ldots).
\end{equation}
{\it Remark 4.2.} If we take the values of $\alpha$ as in Remark
4.1., i.e. $\alpha_k=4k+1$ it is again possible to derive the
inequality for $k\to\infty$. The reasoning is similar as in the
limiting case of inequality (\ref{symmetric}). We take the
logarithm of both sides of Eq. (\ref{iloczyn}) and divide the
inequality by $1-\alpha_k=-4k$ obtaining
\begin{equation}
\frac{\log\Tr\varrho_{B(A)}^{4k+1}}{-4k}\leq\frac{\log
2^{4k}\left[\Tr\varrho\left(\varrho\varrho^{\tau_{A(B)}^{U}}\right)^{2k}\right]}{-4k}.
\end{equation}
It is easy to check that in the limit $k\to\infty$ after omitting
the logarithm we obtain
\begin{equation}\label{limil}
||\varrho_{B}||\geq
2\sqrt{\left|\left|\varrho\varrho^{\tau_{A}^{U}}\right|\right|}.
\end{equation}

{\it Remark 4.3.} In the case when
$[\varrho_A\ot\mathbbm{1}_{d_{B}},\varrho]\neq 0$, it is still
possible to derive certain inequality detecting entanglement,
however, most probably not measurable. To achieve this goal we use
two facts. The first one says that for arbitrary matrices $A$ and
$B$, the following equality
\begin{equation}
\lim_{m\to\infty}\left(e^{\frac{B}{2m}}e^{\frac{A}{m}}e^{\frac{B}{2m}}\right)^{m}=e^{A+B}.
\end{equation}
holds (see Ref. \cite{Bhatia}). Therefore one sees that
\begin{eqnarray}
&&e^{\frac{\alpha+1}{2}\log\varrho+\frac{\alpha-1}{2}\log\varrho^{\tau_{A}}}\\
&&\hspace{0.2in}=\lim_{m\to\infty}
\left(e^{\frac{\alpha+1}{4m}\log\varrho} e^{\frac{\alpha-1}{2m}\log\varrho^{\tau_{A}^{U}}} e^{\frac{\alpha+1}{4m}\log\varrho}\right)^{m}\nonumber\\
&&\hspace{0.2in}=\lim_{m\to\infty}\left(\varrho^{\frac{\alpha+1}{4m}}(\varrho^{\tau_{A}^{U}})^{\frac{\alpha-1}{2m}}
\varrho^{\frac{\alpha+1}{4m}}\right)^{m}.
\end{eqnarray}
and by virtue of the continuity of the trace, we have
\begin{equation}
\Tr\varrho_{B}^{\alpha}\geq \lim_{m\to\infty}\Tr\left[\varrho^{\frac{\alpha+1}{4m}}\left(\varrho^{\tau_{A}^{U}}\right)^{\frac{\alpha-1}{2m}}
\varrho^{\frac{\alpha+1}{4m}}\right]^{m}.
\end{equation}
As the second fact we make use of the inequality \cite{Lieb}:
\begin{equation}\label{Lieb}
\Tr\left[B^{r}(\sqrt{B}A\sqrt{B})^{s}\right]\geq \Tr\left[\left(\Sigma_{\uparrow}(A))^{s}(\Sigma_{\downarrow}(B)\right)^{r+s}\right],
\end{equation}
where $A$ and $B$ are positive $n\times n$ matrices, $r\geq 0$, and $s\geq 1$. Here, $\Sigma_{\uparrow}(A)$ and $\Sigma_{\downarrow}(A)$ are defined as
\begin{equation}
\Sigma_{\uparrow}(A)=
\left(
\begin{array}{cccc}
\sigma_{1} & 0          & \ldots & 0 \\
0          & \sigma_{2} & \ldots & 0 \\
\vdots     & \vdots     & \ddots & \vdots \\
0          & 0          & \ldots & \sigma_{n}
\end{array}
\right)
\end{equation}
and
\begin{equation}
\Sigma_{\downarrow}(A)=
\left(
\begin{array}{cccc}
\sigma_{n} & 0          & \ldots & 0 \\
0          & \sigma_{n-1} & \ldots & 0 \\
\vdots     & \vdots     & \ddots & \vdots \\
0          & 0          & \ldots & \sigma_{1}
\end{array}
\right),
\end{equation}
where $\sigma_{1}\geq \sigma_{2}\geq \ldots \geq \sigma_{n}$ are singular values of $A$, i.e.,
eigenvalues of $|A|=\sqrt{A^{\dagger}A}$.

Substituting $r=0,\;s=m$ and $\sqrt{B}=\varrho^{(\alpha+1)/4m}$, and $A=(\varrho^{\tau_{A}^{U}})^{(\alpha-1)/2m}$
to Eq. (\ref{Lieb}), one arrives at
\begin{equation}
\Tr\varrho_{B}^{\alpha}\geq \lim_{m\to\infty}\Tr\left\{\left[\Sigma_{\uparrow}\big((\varrho^{\tau_{A}^{U}})^{\frac{\alpha-1}{2m}}\big)\right]^{m}
\left[\Sigma_{\downarrow}\big(\varrho^{\frac{\alpha+1}{2m}}\big)\right]^{m}\right\}.
\end{equation}
Moreover, in the case of Hermitian $A$ one has
$\Sigma_{\uparrow}(A^{k})=(\Sigma_{\uparrow}(A))^{k}$, which in
turn allows us to write
\begin{eqnarray}\label{general}
\Tr\varrho_{B}^{\alpha}&\geq& \lim_{m\to\infty}\Tr\left[(\Sigma_{\uparrow}(\varrho^{\tau_{A}^{U}}))^{\frac{\alpha-1}{2}}
(\Sigma_{\downarrow}(\varrho))^{\frac{\alpha+1}{2}}\right]\nonumber\\
&=&\Tr\left[\big(\Sigma_{\uparrow}(\varrho^{\tau_{A}^{U}})\big)^{\frac{\alpha-1}{2}}
\big(\Sigma_{\downarrow}(\varrho)\big)^{\frac{\alpha+1}{2}}\right].
\end{eqnarray}
Since for a density matrix $|\varrho|=\varrho$ the singular values
of $\varrho$ are just its eigenvalues.
\subsection{Operator inequalities.}\label{sub-2.C}
In the proof of Fact 2 we considered an operator inequality given
by Eq. (\ref{nier_op}). As we will see below this operator
inequality is interesting to be analyzed itself. Namely, assuming
that a given $\varrho$ on
$\mathbb{C}^{d_{A}}\otimes\mathbb{C}^{d_{B}}$ is separable and
obeys $[\varrho,\varrho_{A}\ot\mathbbm{1}_{d_{B}}]=0$, then
\begin{equation}\label{1}
(\varrho_{A}\ot\mathbbm{1}_{d_{B}})^{\alpha}\geq\left(\varrho+\varrho^{\tau_{B}^{U}}\right)^{\alpha},
\end{equation}
for natural $\alpha\geq 1$. Equivalently, under the assumption
that $[\varrho,\mathbbm{1}_{d_{A}}\ot\varrho_{B}]=0$, we get the
dual inequality of the form
\begin{equation}\label{2}
(\mathbbm{1}_{d_{A}}\ot\varrho_{B})^{\alpha}\geq
\left(\varrho+\varrho^{\tau_{A}^{U}}\right)^{\alpha}.
\end{equation}
Both inequalities are an immediate consequences of the fact that
if $[A,B]=0$ then $A\geq B\geq 0$ implies $A^{\alpha}\geq
B^{\alpha}$ for real $\alpha >0$.

For states that commute with $\varrho^{\tau^{U}_{B}}$ (i.e.,
$[\varrho,\varrho^{\tau_B^{U}}]=0$), the inequality (\ref{1})
gives rise to the family of inequalities of the form
\begin{equation}\label{4}
(\varrho_{A}\ot\mathbbm{1}_{d_{B}})^{\alpha}\geq\varrho^{
\alpha}+\mathcal{G}_{\alpha}(\varrho),
\end{equation}
where $\mathcal{G}_{\alpha}(\varrho)$ denotes a
linear combination of products of different powers of $\varrho$
and $\varrho^{\tau_B^U}$. The operator $\mathcal{G}_{\alpha}(\varrho)$ obviously
depends on parameter $\alpha$ and is obtained by removing some
positive terms on the RHS of the inequality (\ref{1}).

An example of inequality of the type (\ref{4}) is
\begin{eqnarray}\label{oddcut1}
\hspace{-0.7cm}(\varrho_{A}\ot\mathbbm{1}_{d_{B}})^{\alpha}&\geq&\left(\varrho+\varrho^{\tau_{B}^{U}}\right)^{\alpha}\nonumber\\
\hspace{-0.7cm}&\geq&\frac{1}{2}\left[\left(\varrho+\varrho^{\tau_{B}^{U}}\right)^{\alpha}+
\left(\varrho-\varrho^{\tau_{B}^{U}}\right)^{\alpha}\right],
\end{eqnarray}
where the terms with odd powers of $\varrho^{\tau_B^{U}}$ has been
removed, since for separable states
$\varrho^{m}\big(\varrho^{\tau_{B}^{U}}\big)^{n}\geq 0$ for all
$m,\,n\geq 1$. This is the operator version of the scalar
inequality proposed in Fact 3.
Again, it is enough to assume that
$\big[\varrho,\varrho_{A}\ot\mathbbm{1}\big]$. It may be rewritten
in the form
\begin{equation}\label{oddcut2}
(\varrho_{A}\ot\mathbbm{1}_{d_{B}})^{\alpha}\geq\varrho^{
\alpha}+\sum_{k=1}^{\lfloor \alpha/2\rfloor}\varrho^{
\alpha-2k}\left(\varrho^{\tau_{B}^{U}}\right)^{2k}.
\end{equation}
One should notice, that if the state $\varrho$ has negative
partial time reversal then the removed terms could become
nonpositive and removing them from Eq. (\ref{1}) should make the
inequality more powerful than the Breuer criterion. The comparison
to Breuer criterion and others is presented in the next section.
\subsection{Comparison}\label{sub-2.D}
We shall now compare the scalar inequalities and the operator
inequality introduced in previous paragraphs with the known scalar
and structural separability criteria, paying particular attention
to the entropic inequalities and the criterion formulated by
Breuer \cite{Bcrit}. The large class of states that possesses all
the features necessary to apply the inequalities derived in
previous sections are the rotationally invariant bipartite states
(for some results on separability properties of
$\mathrm{SO}(3)$-invariant states see Refs. \cite{rot,RAJS}). They
have maximally mixed subsystems and their partial time reversal
with respect to arbitrary subsystem does not change the
eigenvectors of a state, so they fulfil the assumption
$[\varrho,\varrho^{\tau_{A(B)}}]=0$. Every bipartite
$\mathrm{SO}(3)$-invariant state with subsystems of spin $j_1$ and
$j_2$ such that $j_1\leq j_2$ can be written in the basis of
projections on eigenspaces of total angular momentum $P_J$, where
$J=|j_1-j_2|,\ldots,j_1+j_2$, i.e.,
\begin{equation}\label{alpha}
\varrho=\sum_{J=|j_{1}-j_{2}|}^{j_{1}+j_{2}}\alpha_{J}P_{J}.
\end{equation}
Normalization is such that $\Tr P_J=1$.

We shall focus our attention on the case of $4\otimes 4$ states
for which entanglement is fully characterized by partial
transposition and Breuer's map, i.e., Breuer criterion in this case
detects all bound entangled states. Each state depends on three
nonnegative parameters $p$, $q$, $r$ such that $0\leq 1-p-q-r\leq
1$ and can be written as
\begin{equation}\label{4x4}
\varrho(p,q,r)=p P_{0}+q P_{1}+r P_{2}+(1-p-q-r) P_{3}.
\end{equation}

\begin{figure}[!hbp]
(a)\includegraphics[width=0.22\textwidth]{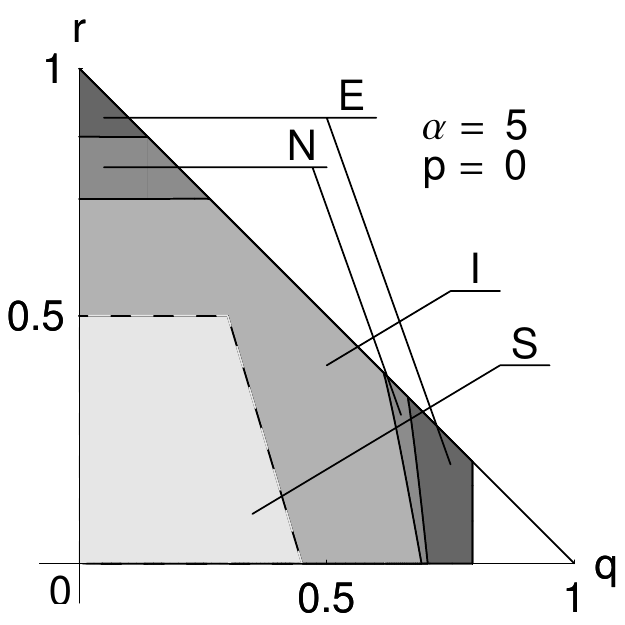}\includegraphics[width=0.22\textwidth]{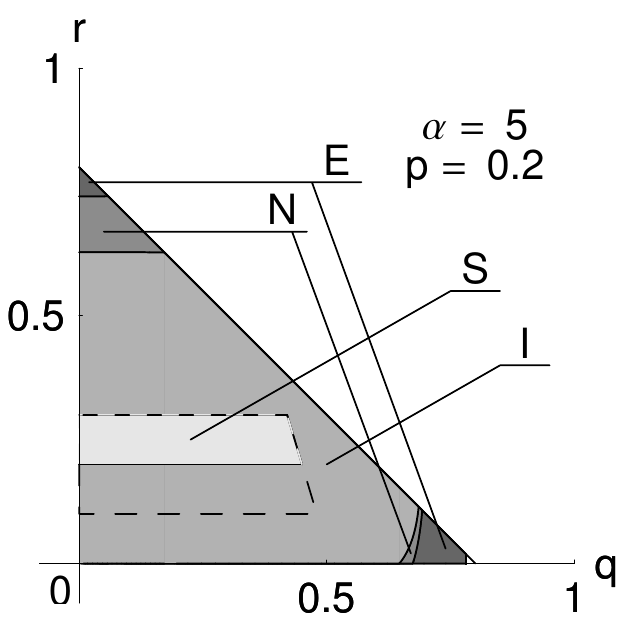}\\
(b)\includegraphics[width=0.22\textwidth]{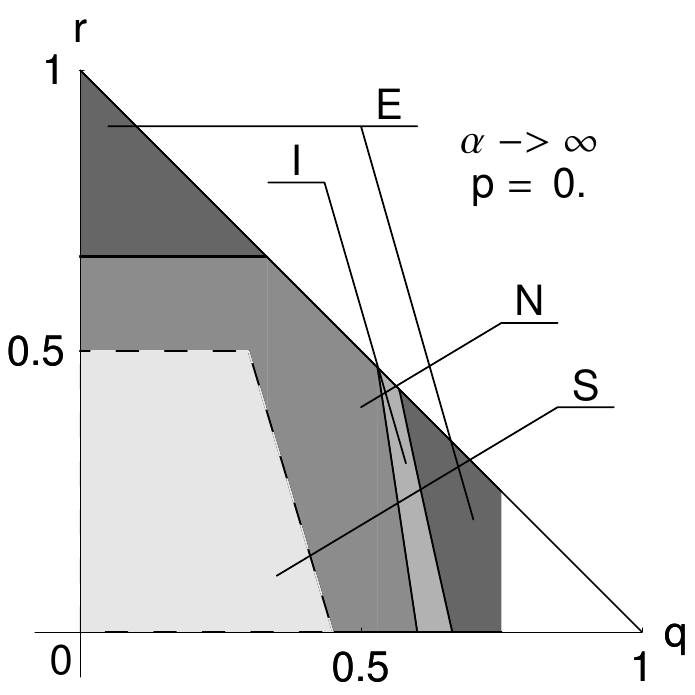}\includegraphics[width=0.22\textwidth]{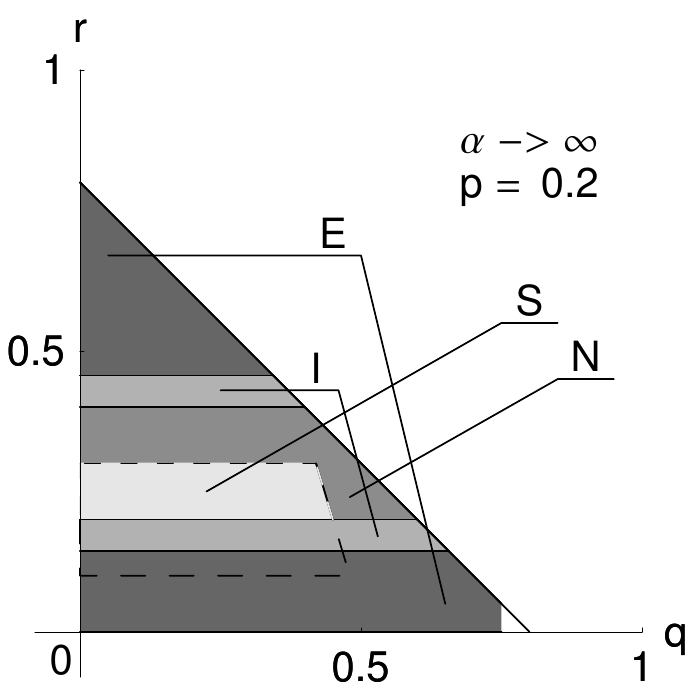}\\
\caption{The comparison of entropic inequalities with these
derived in the present paper for $\alpha=5$, (a) and
$\alpha\to\infty$, (b) and state parameter $p=0$ and $p=0.2$. The
range of parameters $q,r$ that represent a state for given $p$ is
the triangle marked in each picture. The sets for which the
respective inequalities are fulfilled overlap in the area around
the separable states and therefore only some parts of them are
visible in the pictures. To avoid confusion the sets are marked
both with colors and letters. The largest, $E$, the set for which
the $\alpha$-entropic inequality is fulfilled, $N$, inequality
(\ref{symmetric}), $I$, inequality (\ref{iloczyn}) and the
smallest $S$, the set of separable states. The dashed line is the
border of the set of PPT states.}\label{fig4x4ent}
\end{figure}

We start the analysis with comparing the new inequalities
(\ref{symmetric}) and (\ref{iloczyn}) with standard entropic ones.
As shown in Fig. \ref{fig4x4ent} the set of states that fulfil the
entropic inequality is much larger than these for the present
inequalities. Thus the scalar criteria (\ref{symmetric}) and
(\ref{iloczyn}) resulting from the extended reduction map are
indeed much stronger than the entropic ones, since for the same
values of $\alpha$ they detect more entangled states (regions
outside the respective sets). Moreover the significant feature of
the derived inequalities is that they detect PPT entangled states.
However, in the limit $\alpha\to\infty$ the inequality
(\ref{symmetric}) detects all bound entangled states, whereas
inequality (\ref{iloczyn}) only some part of the set.

\begin{figure}[!hbp]
(a)\includegraphics[width=0.22\textwidth]{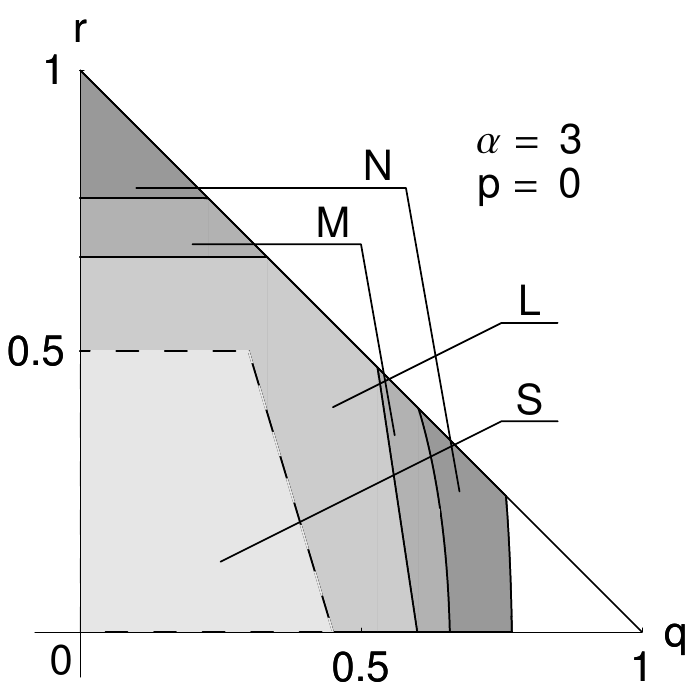}\includegraphics[width=0.22\textwidth]{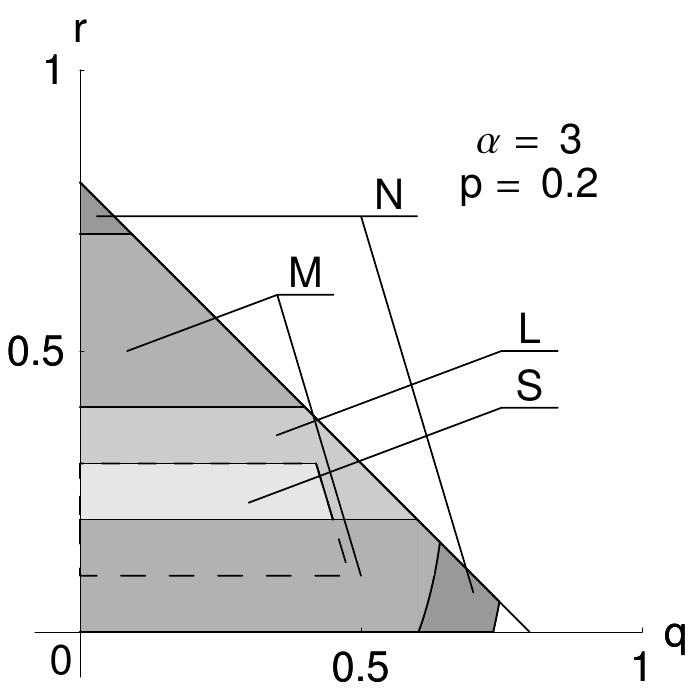}\\
(b)\includegraphics[width=0.22\textwidth]{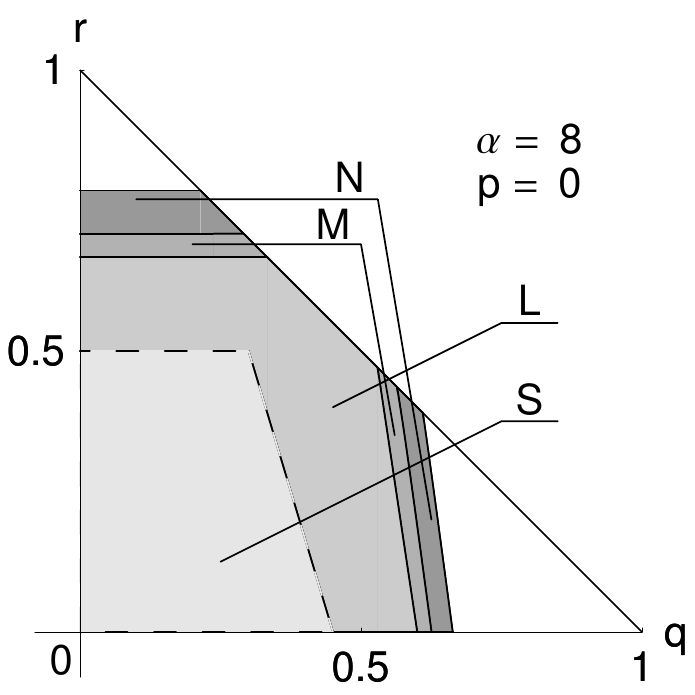}\includegraphics[width=0.22\textwidth]{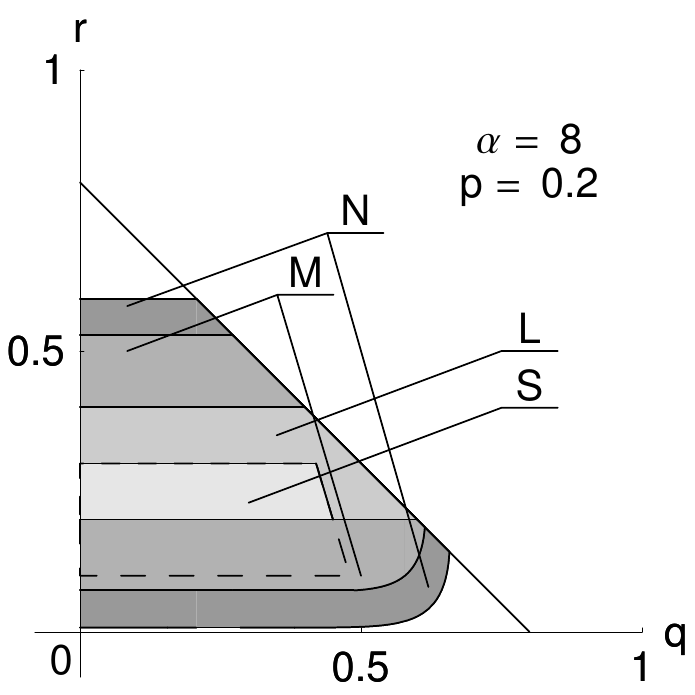}\\
(c)\includegraphics[width=0.22\textwidth]{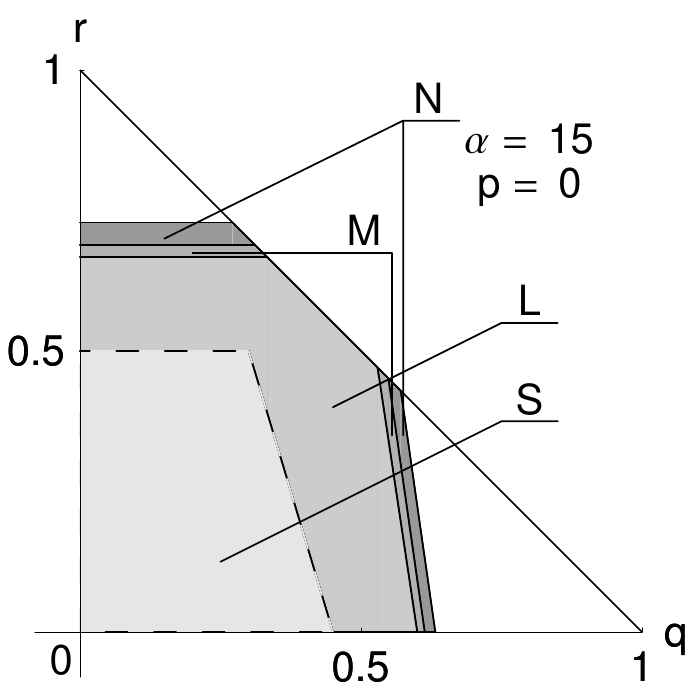}\includegraphics[width=0.22\textwidth]{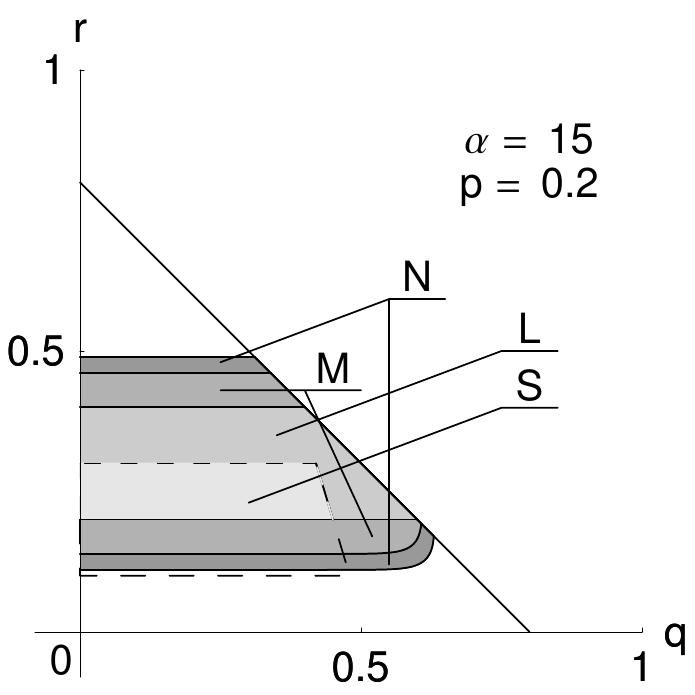}\\
\caption{The comparison of inequalities (\ref{symmetric}) and
(\ref{module}) for $\alpha=3$ (a), $\alpha= 8$ (b), and
$\alpha=15$ (c) and state parameter $p=0$ and $p=0.2$. The set of
parameters $q, r$ which represent a state for given $p$ is the
triangle marked in each figure. Sets for which respective
inequalities are fulfilled overlap in the region surrounding the
set of separable states S (i.e., $S\subset L\subset M\subset N$).
We mark them as follows, $N$, states that fulfil Eq.
(\ref{symmetric}), $M$, Eq. (\ref{module}), $L$, the limit
$\alpha\to\infty$ of Eqs. (\ref{symmetric}) and (\ref{module}).
The dashed line is the border of the set of PPT
states.}\label{fig4x4sym}
\end{figure}

\begin{figure}[!hbp]
(a)\includegraphics[width=0.22\textwidth]{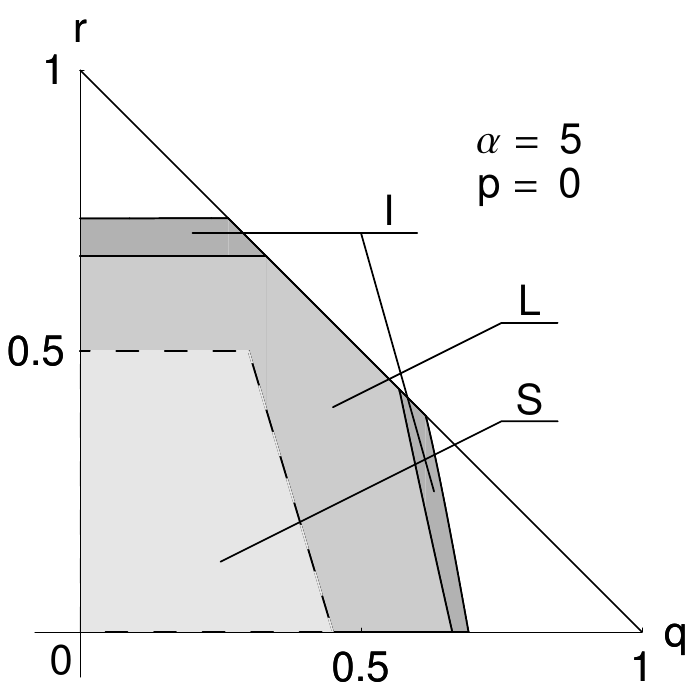}\includegraphics[width=0.22\textwidth]{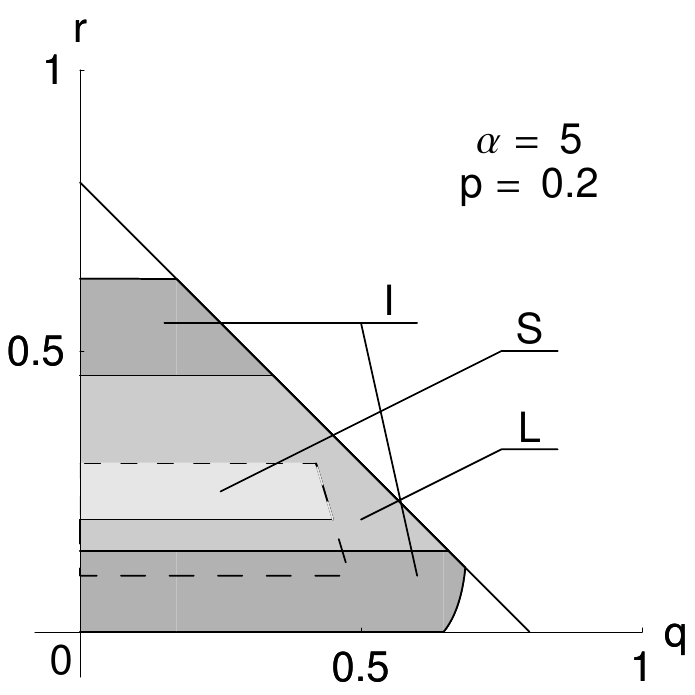}\\
(b)\includegraphics[width=0.22\textwidth]{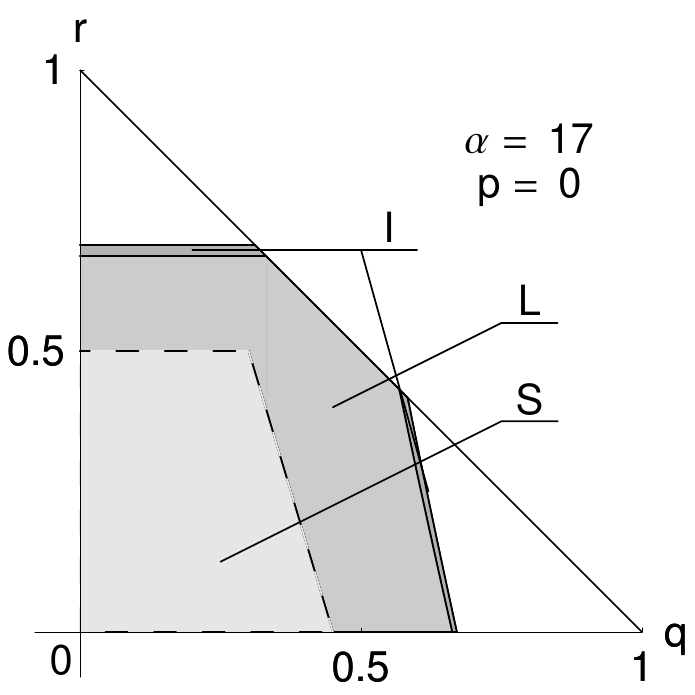}\includegraphics[width=0.22\textwidth]{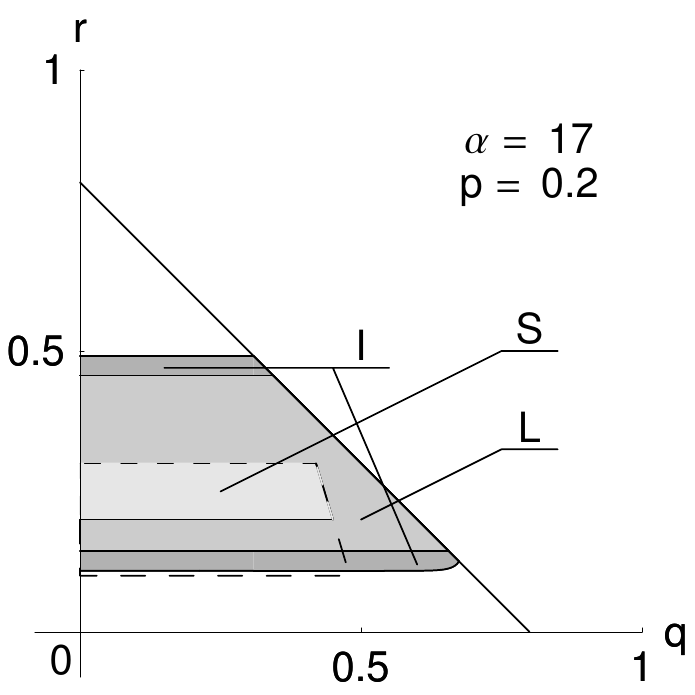}\\
\caption{The set of states that fulfil the Eq. (\ref{iloczyn}) is
marked with $I$, the limiting case $\alpha\to\infty$ of this
inequality is denoted by $L$. Again $S$ denotes the set of
separable states and the dashed line the border of the set of PPT
states. ($S\subset L\subset I$) The triangle is the set of
parameters $q,r$ which represent a state. The figures are made for
two values of state parameter $p=0$ and $p=0.2$ and $\alpha=5$
(a), $\alpha=17$ (b).}\label{fig4x4il}
\end{figure}

In Figs. \ref{fig4x4sym} and \ref{fig4x4il} the effectiveness of
the inequalities proposed in the paper is shown. The comparison of
inequalities (\ref{module}) and (\ref{symmetric}) is made in Fig.
\ref{fig4x4sym}. It can be seen that the second is stronger than
the first one, i.e., detects more entangled states for the same
value of parameter $\alpha$. Comparing the figures in the right
column one can see how the PPT entangled states are detected with
the growth of parameter $\alpha$. In the limit $\alpha\to\infty$
(marked in each figure with $L$) both inequalities detect the
whole set of bound entangled states.

The effectiveness of Eq. (\ref{48}) is shown in Fig.
\ref{fig4x4il}. The set marked with $I$ converges to the one marked by $L$  with
the growing $\alpha$. It should be noticed that even
for relatively small values of $\alpha$ the difference between the
sets $I$ and $L$ is small.

\begin{figure}[!hbp]
(a)\includegraphics[width=0.22\textwidth]{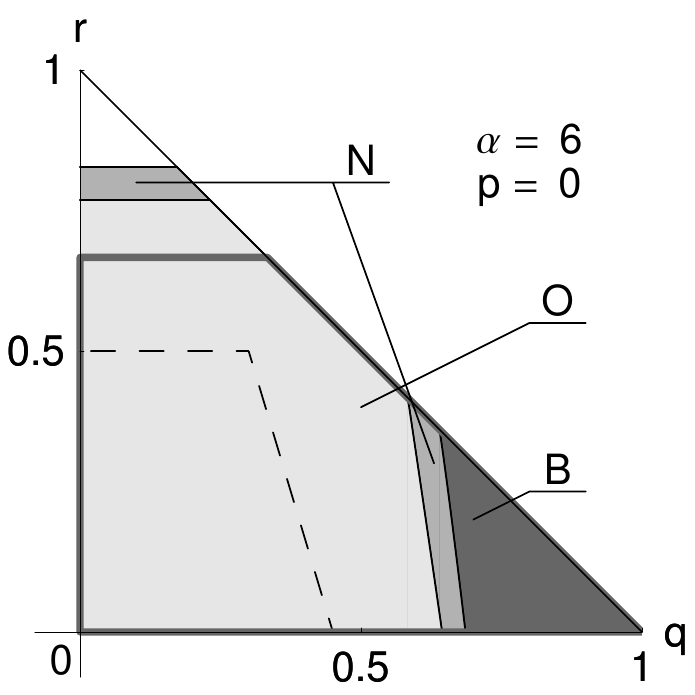}\includegraphics[width=0.22\textwidth]{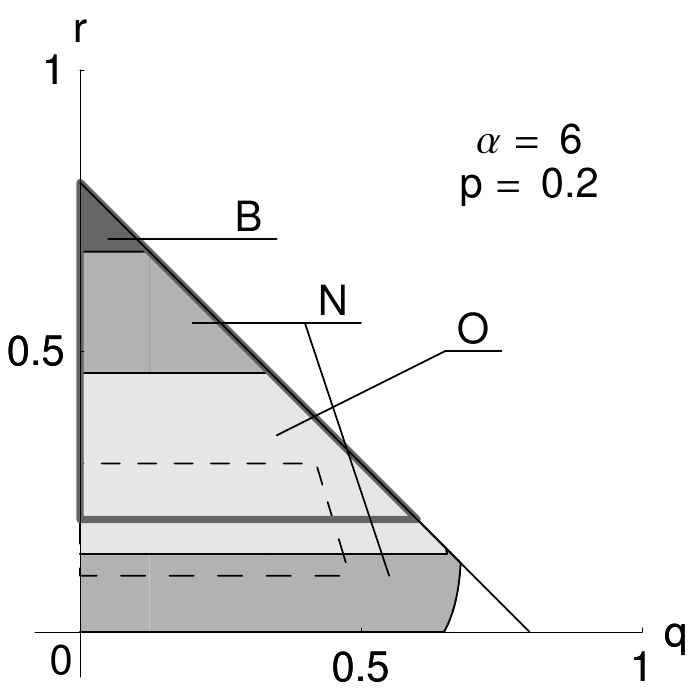}\\
(b)\includegraphics[width=0.22\textwidth]{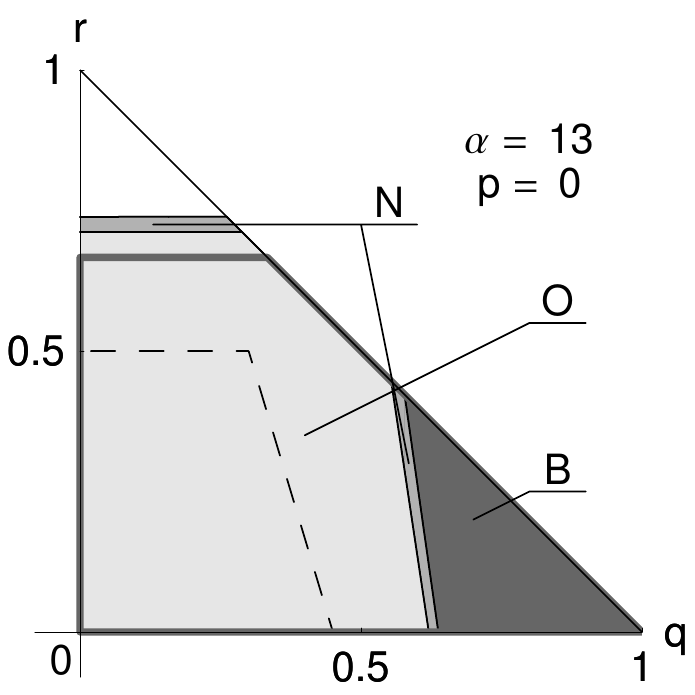}\includegraphics[width=0.22\textwidth]{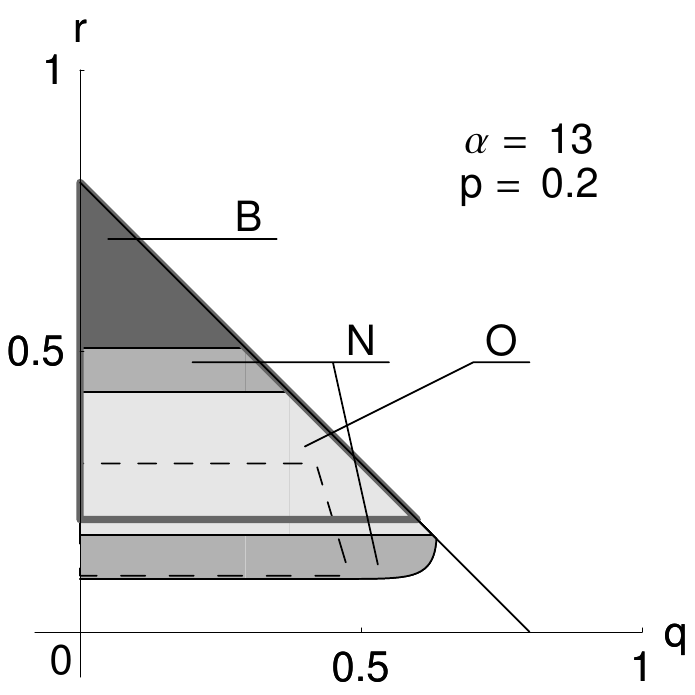}\\
\caption{The comparison of operator inequality (\ref{oddcut1})
with the scalar inequality (\ref{symmetric}) and the Breuer map
criterion for $\alpha=6$ (a) and $\alpha=13$ (b). The set of
parameters $q, r$ which represent a state for $p=0$ and $p= 0.2$
is the triangle marked in each figure. Sets for which respective
inequalities are fulfilled are marked as follows, $B$, the set of
states that remain positive after the action of Breuer's map, the
borders of the set are marked with the thick gray line since the
set is partially covered by other sets. $O$, states that fulfil
Eq. (\ref{oddcut1}), $N$, states that fulfil Eq.
(\ref{symmetric}). The dashed line is the border of the set of PPT
states. The set of separable states is the intersection of $B$ and
the set of PPT states.} \label{fig4x4op}
\end{figure}

In Fig. \ref{fig4x4op} we compare the operator inequality
(\ref{oddcut1}) derived in the previous section with the positive
map criterion proposed by Breuer. The figures contain also the
scalar inequality (\ref{symmetric}) since it may be considered as
a scalar analog of (\ref{oddcut1}). It is clearly seen that the
operator inequality (\ref{oddcut1}), though arising from the
Breuer's map, detects some entanglement where the Breuer's map
fails. The scalar inequality is weaker than the operator one,
however, in the limit $\alpha\to\infty$ both criteria are
equivalent for this class of states.

%
\section{Multi-copy entanglement witnesses.}\label{witness}
\setcounter{equation}{0}
Here we discuss the applicability of just introduced scalar
inequalities for construction of multicopy entanglement witnesses.
One knows that measuring such observables may provide more
information about entanglement of a given $\varrho$ than witnesses
defined on a single copy. In particular, recently two-copy
entanglement witnesses were shown to be a lower bound for
concurrence of $\varrho$ \cite{Mintert}.

First, the notions of the $n$-copy observable and $n$-copy
entanglement witness were proposed in Ref. \cite{PH3}. The latter
are Hermitian operators $\mathscr{W}^{(n)}$ such that their mean
value on $n$ copies of any separable state $\varrho$ is positive
and there exists an entangled state for which this mean value is
negative. An example of such operator unambiguously determining
whether the state is entangled was provided in Ref. \cite{RAPHMD}
for any two-qubit state and in Ref. \cite{RAJS} for $2\ot d$
rotationally invariant states with odd $d$.

Below we will show how the scalar inequalities considered in the
present paper can be reformulated in terms of a single collective
witness.
First we present the general multicopy approach to entanglement
tests in both scalar (based on witnesses) and structural (based on
maps) scenarios. Consider the scalar inequalities provided in
Secs. \ref{sub-2.A} and \ref{sub-2.B}. They are all of the form
\begin{equation}\label{scalar-inequality}
\Tr\left(\sum_{i=1}^{m}\mu_i\prod_{j=1}^{\alpha}\Theta_{ij}(\varrho)\right)\geq
0
\end{equation}
for some linear maps $\Theta_{ij}$ that preserve Hermiticity and
coefficients $\mu_i\in\mathbb{R}$. For instance, the inequality
given by Eq. (\ref{symmetric}) has this form if $m=3$ and
\begin{eqnarray}\label{maps1}
\Theta_{1j}&=&\Tr_{B} \qquad\quad\, (j=1,\ldots,\alpha),\nonumber\\
\Theta_{21}&=&\Theta_{31}=I,\quad\nonumber\\
\Theta_{2j}&=&I+\tau_{B}^{U}\qquad (j=2,\ldots,\alpha),\nonumber\\
\Theta_{3j}&=&I-\tau_{B}^{U}\qquad (j=2,\ldots,\alpha),
\end{eqnarray}
$\mu_1=1,\, \mu_2=\mu_3=-1/2$, $I$ is an identity map, and
$\Tr_{B}$ denotes the partial trace over the second subsystem.

Now, assuming that $\Tr[\Pi_{j}\Theta_{ij}(\varrho)]\in\mathbb{R}$
for all $i$, we provide multicopy entanglement witnesses that
follow from the above scalar inequalities and go beyond these
provided for entropic inequalities \cite{PH3}. First, let us
denote by $\mathscr{V}^{(n)}$ the $n$-copy swap operator
\begin{equation}\label{6}
\mathscr{V}^{(n)}\ket{\Phi_{1}}\ket{\Phi_{2}}\ldots\ket{\Phi_{n}}=
\ket{\Phi_{n}}\ket{\Phi_{1}}\ldots\ket{\Phi_{n-1}},
\end{equation}
which is a straightforward multipartite generalization of
$\mathscr{V}^{(2)}$ introduced in Sec. \ref{sub-2.A}. It has the
property that
\begin{equation}\label{V}
\Tr(\mathscr{V}^{(n)}\varrho_{1}\ot\ldots\ot\varrho_{n})=\Tr(\varrho_{1}\ldots\varrho_{n}).
\end{equation}
However, one should notice that
\begin{equation}
\Tr(\mathscr{V}^{(n)\dagger}\varrho_{1}\ot\ldots\ot\varrho_{n})=\Tr(\varrho_{n}\ldots\varrho_{1}),
\end{equation}
which is not the same as in Eq. (\ref{V}). The equivalence between
these two formulas exist only if both traces are real.

One can see that $\mathscr{V}^{(n)}$ is not a Hermitian operator and as such it cannot be treated as an observable.
However, instead of $\mathscr{V}^{(n)}$ one may consider its Hermitian counterpart
\begin{equation}
\tilde{\mathscr{V}}^{(n)}=\frac{1}{2}\left(\mathscr{V}^{(n)}+\mathscr{V}^{(n)\dagger}\right),
\end{equation}
of which the mean value on $n$ copies of the state $\varrho$ gives exactly $\Tr\varrho^{n}$.
Now, to take into account the maps in the formula (\ref{scalar-inequality}), we use approach
exploited already in case of positive maps method \cite{PHRAMD} and
define the following collective witness:
\begin{equation}
\mathscr{W}^{(\alpha)}=\sum_{i=1}^{m}\mu_i\left(\bigotimes_{j=1}^{\alpha}\Theta_{ij}^{\dagger}\right)\left(\tilde{\mathscr{V}}^{(\alpha)}\right)
\end{equation}
which is Hermitian by the construction. Here by $\Theta^{\dagger}$
we denote a dual map of $\Theta$, i.e., the map obeying
$\Tr[X\Theta(Y)]=\Tr[\Theta^{\dagger}(X)Y]$ for all matrices $X$
and $Y$.

Then the collective witness inequality that is equivalent to Eq.
(\ref{scalar-inequality}) is of the form
\begin{equation}
\Tr(\mathscr{W}^{(\alpha)} \varrho^{\otimes \alpha})\geq 0.
\end{equation}
As illustrative examples we consider witnesses following from
inequalities given by Eq. (\ref{symmetric}) and the ones given by
Eq. (\ref{48}). In the first case one needs to take dual maps
$\Theta^{\dagger}_{ij}$ $(i=1,2,3,\; j=1,\ldots,\alpha)$ of the
ones defined by Eq. (\ref{maps1}). In the second case one takes
$\Theta_{1j}$ for $j=1,\ldots,4k+1$ as defined in the previous
case and
\begin{eqnarray}
&&\Theta_{2j}=I \quad\quad (j=1,\ldots,2k+1),\nonumber\\
&&\Theta_{2j}=\tau_{A(B)}^{U}\quad (j=2k+2,\ldots,4k+1),\nonumber\\
&&\mu_1=1,\quad\mu_2=-2^{4k}.
\end{eqnarray}

Now we come back to operator inequalities of the type proposed in
the Sec. \ref{sub-2.C}. They are all of the form
\begin{equation}\label{generalmap}
\sum_{i=1}^{m}\mu_i\prod_{j=1}^{\alpha}\Theta_{ij} (\varrho)\geq 0,
\end{equation}
where again $\Theta_{ij}$ are Hermiticity-preserving linear maps.
Here we shall proceed in a slightly different way to highlight the
analogy to positive maps separability condition. Namely, we can
define the linear, map $\Lambda^{(n)}:{\cal H}_{AB}^{\otimes n}
\rightarrow {\cal H}_{AB}$ by the formula
\begin{equation}
\Lambda^{(n)}(\cdot)= \hspace{-0.3cm}\sum_{k,m,i_{1},\ldots,i_{n}}
\hspace{-0.3cm} P_{km} \Tr[P_{ki_{1}} \otimes P_{i_{1}i_{2}}
\otimes P_{i_{2}i_{3}} \otimes \ldots \otimes P_{i_{n}m}(\cdot)]
\label{(n)}
\end{equation}
with $P_{ij}=|i \rangle \langle j|$. The above map
satisfies $\Lambda^{(\alpha)}(A_{1} \otimes A_{2} \otimes \ldots \otimes A_{\alpha})=A_{1}A_{2}\ldots A_{\alpha}$
for any operators $A_{i}$. Using the above map we can define
the map
\begin{equation}
\Theta^{(\alpha)}=\Lambda^{(\alpha)} \circ
\sum_{i=1}^{m}\mu_i\bigotimes_{j=1}^{n}\Theta_{ij}
\end{equation}
and then the operator inequality (\ref{generalmap}) looks as follows
\begin{equation}
\Theta^{(\alpha)}(\varrho^{\otimes \alpha})\geq 0.
\end{equation}
Since this inequality is satisfied iff
$\bra{\Psi} \Theta^{(\alpha)}(\varrho^{\otimes \alpha})\ket{\Psi} \geq 0$ for
any vector $\ket{\Psi}$, we can immediately provide infinite set of
$n$-copy entanglement witnesses
\begin{equation}
\mathscr{W}_{\Psi}^{(\alpha)} \equiv \Theta^{(\alpha) \dagger}(\proj{\Psi}).
\end{equation}
%
%
\section{Special inequality with the reflection map and
its representation in terms of experimental quantities}\label{IV}
\setcounter{equation}{0}

\subsection{Quadratic inequality based on reflection}\label{IV.A}
\begin{figure}
(a)\includegraphics[width=0.47\textwidth]{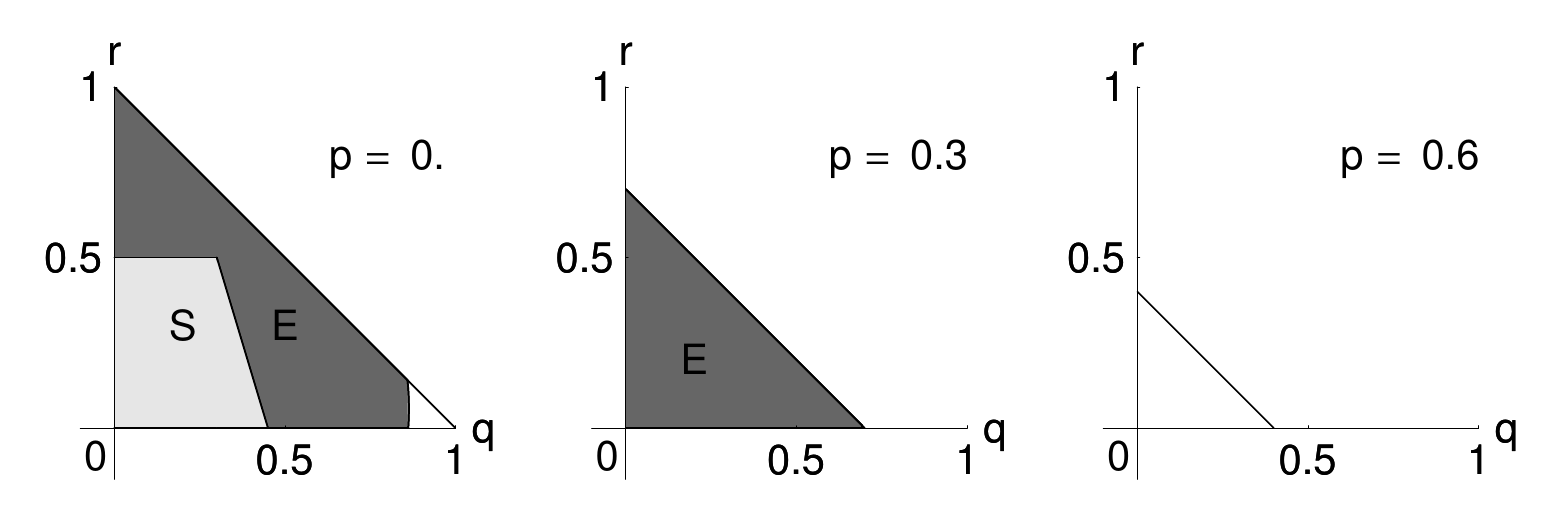}\\
(b)\includegraphics[width=0.47\textwidth]{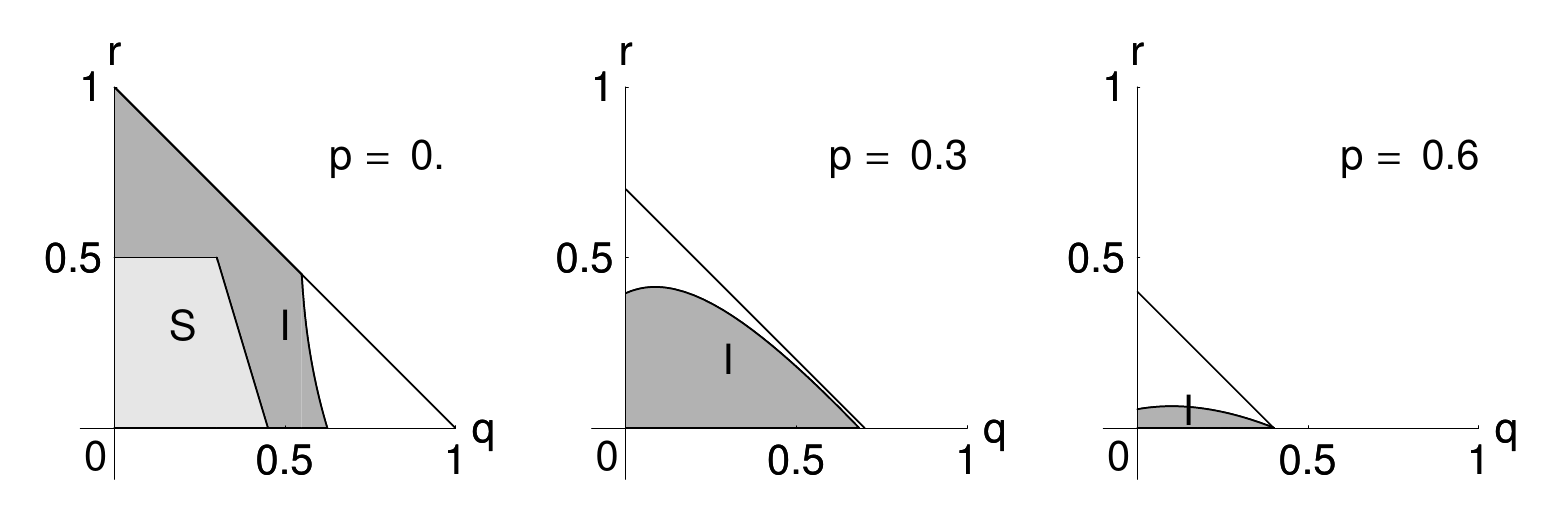}\\
  \caption{The comparison of the entropic inequalities (\ref{entropic}) for $\alpha=2$ and
  $\Tr(\varrho\varrho^{\tau_{A(B)}})\geq{0}$ for $4\otimes 4$ rotationally
  invariant states (\ref{4x4}). In all figures the light gray area marked with $S$ represents separable
  states. (a) The set of states that fulfill the entropic inequality
  $\Tr\varrho_{A(B)}^{2}-\Tr\varrho^{2}\geq0$ (dark gray region marked with $E$). (b) The set of states that
  fulfill the inequality $\Tr(\varrho\varrho^{\tau_{A(B)}})\geq 0$ (dark gray region marked with $I$).
  The range of parameters $q$ and $r$ which represent the state for $p=0$, $p=0.3$, and $p=0.6$ is the triangle
  marked in all the figures.}\label{ent_rrt}
\end{figure}

Following the PPT test it is immediate to see that the following
inequality is satisfied for any separable state
\begin{equation}
\Tr\left(\varrho \varrho^{\tau^{U}_{A(B)}}\right)\geq 0, \label{ineq1}
\end{equation}
The above condition is related to the entropic inequality
(\ref{entropic0}) by Eq. (\ref{1tr}) (both taken with $\alpha=2$)
\begin{equation}
\Tr\varrho_{A(B)}^{2}-\Tr\varrho^{2}
\geq \Tr\left(\varrho\varrho^{\tau_{B(A)}^{U}}\right).
\end{equation}
So whenever this inequality is fulfilled (this is the case for
some entangled states) Eq. (\ref{ineq1}) may be violated
independently of respective entropic inequality. The effectiveness
of inequalities in case of rotationally invariant states
considered in Sec. $\ref{sub-2.D}$ is presented in Fig.
\ref{ent_rrt}.
Below we shall consider experimental detectability of the inequality (\ref{ineq1})
in case of the bipartite systems simulated by multiqubit ones.

\subsection{Experimental detection of  $\varrho \varrho^{\tau}$
for multiqubit systems}\label{IV.B}
Consider an arbitrary state of $n$ qubits $\varrho_{A_{1}\ldots
A_{n}}$. By the map $\tau_{i_0}$ we shall denote the
reflection of $i_0$-th qubit on a Bloch sphere, i.e.,
\begin{equation}
\tau_{i_0}(\varrho_{A_{1},\ldots,A_{n}})=\mathbbm{1}_{n\setminus
i_{0}}\otimes \sigma_{y}^{A_{i_{0}}}\varrho_{A_{1},\ldots,
A_{n}}^{\Gamma_{i_{0}}}\mathbbm{1}_{n\setminus i_{0}}\otimes
\sigma_{y}^{A_{i_{0}}},
\end{equation}
where $\mathbbm{1}_{n\setminus i_{0}}$ means an identity acting on
all parties excluding the $i_{0}$th one and $\Gamma_{i_{0}}$
denotes the partial transposition taken with respect to the
$i_{0}$th particle. Since for any $i,j$ the maps $\tau_{i}$,
$\tau_{j}$ commute, it makes sense to define for any set of
increasing indices \cite{footnote3} $I'=\{i_{1},\ldots,i_{k} \}$
the map
\begin{eqnarray}
\tau_{i_1,\ldots,i_k}(\cdot) =\tau_{i_1} \circ
\ldots \circ \tau_{i_k}(\cdot)= (\cdot)^{\tau_{I'}}.
\end{eqnarray}
The latter notation will be used subsequently.
Now we are interested in general in measuring the
following quantity
\begin{equation}
\Tr(\varrho\varrho^{\tau_{I'}}).
\label{ineq3}
\end{equation}
It may be a little bit surprising that for two-qubit photon
polarization state $\varrho$ the above quantity can be measured
with {\it virtually the same setup} as in Ref. \cite{Bovino}. It
consists of two sources of pairs of photons and performs joint
measurements on their polarization degrees of freedom. First
source emits photons $A$ and $B$, the second one $A'$ and $B'$.
Then the photons $A$ and $A'$ ($B$ and $B'$, respectively) meet at
the beam splitter on Alice (Bob) side. Further, one measures
whether they went out of the beam splitter together (coalescence)
or not (anticoalescence) which formally corresponds to projection
of two photon polarization state onto symmetric or antisymmetric
(singlet) subspace. Since this happens on both Alice and Bob side
the setup finally allows one to measure joint probabilities
(corresponding to coalescence-coalescence,
coalescence-anticoalescence, anticoalescence-coalescence and
anticoalescence-anticoalescence). For $n$ qubits the direct
generalization of the latter experiment (mathematics of which was
considered in Ref. \cite{AlvesBosons}) also happens to work. The
essential difference lies in the way, in which one has to combine
the probabilities that come out from the experiment. Let us derive
the probabilistic formula for the quantity (\ref{ineq1}). Consider
the following collective two-copy multiqubit entanglement witness
\begin{eqnarray}
W^{(2)}_{I'}= \bigotimes_{i\in I \setminus I'}
\mathscr{V}^{(2)}_{A_i A'_i} \bigotimes_{k \in I'} P^{(-)}_{A_k
A'_k}. \label{observable}
\end{eqnarray}
Here $\mathscr{V}^{(2)}$ is a two-partite swap operator as defined
in Eq. (\ref{6}) and
$P^{(-)}=(1/2)(\mathbbm{1}-\mathscr{V}^{(2)})$ is the
antysymmetric projector, which is essentially the projector onto
the singlet state $\ket{\psi_{-}}$. Let us recall that the subset
of indices $I'$ enumerates the qubits, on which the reflection in
the second copy is to be performed. We have explicitly put the
dependence of the observable on that set of indices.

After a little bit of algebra we get that our quantity
(\ref{ineq1}) is reproduced as collective mean value of the
observable
\begin{equation}
\meanb{W^{(2)}_{I'}} \equiv \Tr(W^{(2)}_{I'} \varrho \otimes
\varrho)= \Tr(\varrho\varrho^{\tau_{I'}}).
\end{equation}
Note that this quantity is closely related to entropic
inequalities on multipartite qubits considered in Ref.
\cite{AlvesBosons} (which are natural generalizations of original
entropic inequalities \cite{HHHEntr}) as well as to multipartite
concurrences \cite{MintertMulti} or other state functions based on
nonlinear operations \cite{Osterloh}.

Now there is a question how to measure the mean value of
$W^{(2)}_{I'}$. Consider the generalization of the scheme
\cite{Bovino} in a spirit of Ref. \cite{AlvesBosons}. In this case
on each pair of qubits $A_{i}A'_{i}$ $(i=1,\ldots,n)$ one performs
the measurement projecting onto one of two projectors, i.e.,
symmetric or antisymmetric one $P^{(\pm)}=(1/2)(\mathbbm{1}\pm
\mathscr{V}^{(2)})$. For simplicity  we shall denote the symmetric
and antysymmetric projector by slightly different notations
$P^{(0)}$ and $P^{(1)}$, respectively (note that the index is even
or odd when the symmetry is even or odd).

With help of this notation let us denote the joint
probabilities resulting in the experiment by
\begin{equation}
p(s_1,\ldots,s_n)= \Tr
\left(\bigotimes_{j=1}^{n}P^{(s_{j})}_{A_jA'_j} \varrho_{A_1\ldots
A_n} \otimes \varrho_{A'_1\ldots A'_n}\right),
\end{equation}
where $s_{i}\in \{0,1\}$. Now we derive the mean value of the
observable (\ref{observable}). Let us define the characteristic
function $\chi_{I'}$ of a set of indices $I'$ in a standard way,
i.e., $\chi_{I'}(i)= 1$ if $i\in I'$ and zero otherwise. Let us
also introduce the function $\tilde{\chi}(s_{i})\equiv
\chi_{I'}(i)\delta_{s_i,1}+\chi_{I\setminus I'}(i)$. Then the mean
value of the observable (\ref{observable}) is
\begin{eqnarray}
\meanb{W^{(2)}_{I'}}&=&\sum_{s_1,\ldots,s_n} (-1)^{\sum_{i}s_i
\chi_{I\setminus
I'}(i)}\left[\tilde{\chi}(s_{1})\ldots\tilde{\chi}(s_{n})\right]\nonumber\\
&&\times p(s_{1},\ldots,s_{n}).
\end{eqnarray}

The above complicated-looking formula has a very elementary
interpretation. In fact we are summing only over such $s_{i}$ that
have index $i \notin I'$ and only they contribute to the ''phase''
in the sum. All the indices $s_{i}$ with $i \in I'$ are put to be
one all the time. This can be easily seen in the following
examples.

{\it Example 1.} Consider three qubits $(n=3)$ with the last one
reflected $(I'=\{3\})$. Then the last index in the probability is
fixed to be one while the others are counted. This gives
\begin{eqnarray}
&&\hspace{-0.5cm}\meanb{W^{(2)}_{\{ 3 \}}}=\sum_{i,j=0}^{1} (-1)^{i+j} p(i,j,1)= p(0,0,1)-p(0,1,1)\nonumber \\
&&\hspace{1.4cm}-p(1,0,1)+p(1,1,1).
\end{eqnarray}
Now we introduce further examples that will have important
interpretation in the context of bipartite systems.

{\it Example 2.} Consider again three qubits $(n=3)$ with last two
reflected $(I'=\{2,3\})$. This gives the very easy formula
\begin{equation}
\meanb{W^{(2)}_{\{2,3 \}}} = \sum_{i=0}^{1} (-1)^{i} p(i,1,1)=
p(0,1,1)-p(1,1,1).
\end{equation}

{\it Example 3.} Here we shall focus on four qubits $(n=4)$ and
reflect the last two $(I'=\{3,4\})$. The corresponding formula is
\begin{eqnarray}\label{44}
&&\meanb{W^{(2)}_{\{3,4 \}}}=
\sum_{i,j=0}^{1} (-1)^{i+j}p(i,j,1,1)= p(0,0,1,1) \nonumber \\
&& -p(0,1,1,1)-p(1,0,1,1)+p(1,1,1,1).
\end{eqnarray}
%
\subsection{Application to bipartite systems of higher dimensions}\label{IV.C}
Higher-dimensional bipartite systems, i.e., $d_A \otimes d_B$ with
$d_A d_B>6$ behave in general much different than low-dimensional
ones ($MN\leq 6$). In Sec. \ref{IV.A} we have seen this from
comparison of two scalar inequalities. In the last section we have
considered abstract problem of detection of some quantity for
multiqubit systems. To see how it can work for bipartite one let
us suppose that we are interested in experimental demonstration of
the inequality (\ref{ineq1}) for higher dimensional bipartite
system $AB$. With three-qubit state, say, in polarization
generated with a single source we can simulate $2 \otimes 4$
system interpreting the first qubit as a subsystem $A$ and the
second two as a joint subsystem $B$. Then the Example 2 above
gives immediately an experimental realization of the inequality
(\ref{ineq1}). The particular importance of the inequality is that
it involves only two probabilities and as such should be
experimentally more feasible than the other ones.

Another important example is the one corresponding to $4 \otimes
4$ system. This is because reflection map plays an important
role in the indecomposable Breuer map. Any four qubit state
can be interpreted in this way and then the formula (\ref{44})
serves as an experimental simulation of the bipartite test
(\ref{ineq1}).

Finally note that for two qubits the left-hand side of the
analyzed inequality is $p(0,1)-p(1,1)$ which is just the
difference of anticoalescence and coalescence terms in experiment
\cite{Bovino}. In other words the left hand side of Eq.
(\ref{ineq1}) can be easily calculated basing on experimental
results of Ref. \cite{Bovino}. It amounts to $\Tr(\varrho
\varrho^{\tau})=p(0,1)-p(1,1)=-0.2330 \pm 0.016 < 0$ which clearly
violates the inequality. In this case we have a kind of (undirect)
experimental illustration of the presented approach. It must be
stressed, though, that in this case (as in all $2 \otimes d$ cases
with reflection performed on the smaller system, which includes
Example 1 in Sec. \ref{IV.B}) the analyzed inequality is fully
equivalent to entropic inequality (\ref{entropic0}) taken with
$\alpha=2$. This is not, however, the case for Examples 2 and 3
(see Sec. \ref{IV.B}).

\section{Conclusions}
\setcounter{equation}{0}

The so called entropic inequalities are one of the best known
scalar separability criteria. However, being a direct consequence
of the reduction map, they are not useful in detecting bound
entanglement.

In the present paper we go beyond the reduction map and derive
much stronger entropic-like inequalities from the recently
introduced extended reduction criterion \cite{Bcrit, Bcrit2,Hall}.
The commutativity conditions make the inequalities applicable to a
particular, however large, class of states including the states
isomorphic to quantum channels.
The comparison to known criteria, i.e., G\"uhne-Lewenstein
inequalities \cite{Guhne} for two-qubit states, entropic
inequalities
\cite{RPHPLA,HHHEntr,CAG,MRHPRA,Terhal3,VollbrechtWolf}, and
Breuer witness  \cite{Bcrit} for $4\ot 4$ rotationally invariant
bipartite states shows the effectiveness of the new inequalities
in detection of both distillable and bound entanglement. It should
be emphasized that due to the assumption about positive partial
transposition used in derivation of the inequalities they detect
some NPT entanglement in regions where the Breuer witness fails.
This is especially apparent if one takes the limit
$\alpha\to\infty$ from the inequalities (\ref{symmetric}) and
(\ref{iloczyn}). In case of the discussed
$\mathrm{SO}(3)$-invariant states the obtained separability
criteria (\ref{module}) and (\ref{symmetric}) not only detect
bound entangled states equivalently to Breuer witness but also
almost the whole region of NPT states. However, if one wants to
detect PPT entanglement effectively (i.e., using fewer copies of a
state) it is better to apply the inequality (\ref{1tr}).

By virtue of the recent results the derived inequalities provide a
simple way to construct a many-copy (collective) entanglement
witnesses. As discussed the inequalities may also be strengthen
due to the fact that for separable states
$\tau_{A(B)}^{U}(\varrho)\geq 0$. Therefore when deriving
inequalities one may consider $\big|\tau_{A(B)}^{U}(\varrho)\big|$
instead of $\tau_{A(B)}^{U}(\varrho)$. However, these approach is,
to our knowledge, not useful in experimental realizations.

On the other hand the proposed collective entanglement witnesses
seem to be experimentally feasible at least for low values of
parameter $\alpha$ which corresponds to number of copies of a
state measured at a time. It is interesting that as a by-product
of the above analysis we have come across a simple inequality
which can be naturally implemented using the known experimental
schemes on photon polarizations. In particular the results of the
experiment on the usual two-entropy \cite{Bovino} can be easily
reinterpreted in terms of this inequality.

Though the effectiveness of the inequalities presented in the
paper, to our knowledge, one may derive them only in special
cases, namely, assuming some commutation relations. Therefore the
presented results leave much place for further investigation.
Then it seems interesting to investigate the dependence of
efficiency of detecting entanglement or bound entanglement on the
matrix $U$ used in construction of the map $\tau^{U}$. It would be
also desirable to derive an inequality similar to Eq.
(\ref{iloczyn}) without the assumption of commutation of $\varrho$
and $\tau_{A(B)}^{U}(\varrho)$, and stronger than Eq.
(\ref{general}). Finally, it seems interesting to pose the general
question, which states satisfy the assumed commutation relations
and what can we say about entanglement of a given density matrix
$\varrho$ knowing that it obeys them. We leave these questions as
open problems for further research.

\acknowledgments

Discussions with M. Demianowicz are acknowledged. The work was
supported by the Polish Ministry of Science and Higher Education
under the Grant No. 1 P03B 095 29 and EU Integrated Project SCALA.
R. A. also gratefully acknowledges the support of Foundation for
Polish Science.


\begin{thebibliography}{0}
\bibitem{footnote1}This could be proved using the fact that
positive matrices are always Hermitian and have positive
eigenvalues. Then the spectral decompositions
$A=\sum_{i}\lambda_{i}^{A}\proj{\phi_{i}^{A}}$ and
$B=\sum_{i}\lambda_{i}^{B}\proj{\phi_{i}^{B}}$ lead to
%
\begin{equation}
\Tr AB=\sum_{i,j}\lambda_{i}^{A}\lambda_{j}^{B}\big|\braket{\phi_{i}^{A}}{\phi_{j}^{B}}\big|^{2}\geq 0.\nonumber
\end{equation}
\bibitem{footnote2}We assume here $\alpha$ to be natural number since in the case of entangled
$\varrho$ the operator $\varrho+\tau_{A(B)}^{U}(\varrho)$ may not
be positive.
\bibitem{footnote3} In fact the ordering is not required and is put
here only for convinience.
\end{thebibliography}
\end{document}